%% file: main_merged.tex
\documentclass[iop,apj,twocolappendix,numberedappendix]{emulateapj}
\usepackage[utf8]{inputenc}
\usepackage{amsfonts}
\usepackage{amsmath}
\usepackage{mathrsfs}
\usepackage{amssymb}
\usepackage{textgreek}
\usepackage[nolist,nohyperlinks,printonlyused]{acronym}
\usepackage[breaklinks,colorlinks,citecolor=blue,unicode]{hyperref}
\usepackage[english]{babel}
\usepackage{savesym}
\usepackage{verbatim}
\usepackage{printlen}
\savesymbol{tablenum}
\usepackage{siunitx}
\restoresymbol{SIX}{tablenum}
\received{09/20/2017}
\revised{11/07/2017}
\accepted{11/15/2017}

\bibpunct{(}{)}{;}{a}{}{,}

\newcommand{\HI}{\ion{H}{1}}

\newcommand{\HeI}{\ion{He}{1}}
\newcommand{\HeII}{\ion{He}{2}}

\newcommand{\SiIII}{\ion{Si}{3}}

\newcommand{\mathHI}{{\mbox{\scriptsize \HI}}}

\newcommand{\mathSiIII}{{\mbox{\scriptsize \SiIII}}}

\DeclareSIUnit \parsec{pc}
\DeclareSIUnit \h {\ensuremath{\mathit{h}}}
\DeclareSIUnit \com{\ensuremath{c\!}}
\DeclareSIUnit\year{yr}

\newcommand{\kms}{\kilo\meter\per\second}
\newcommand{\skm}{\second\per\kilo\meter}
\newcommand{\Mpc}{\mega\parsec}
\newcommand{\kpc}{\kilo\parsec}
\newcommand{\hinvMpc}{\per\h\Mpc}
\newcommand{\hinvkpc}{\per\h\kpc}
\newcommand{\ckpc}{\com\kpc}

\newcommand{\invisible}[1]{}

\defcitealias{palanque-delabrouille2013onedimensionalLy}{PD+13}

\def\equationautorefname~#1\null{eqn.~(#1)\null}

\begin{document}

\title[Power Spectrum of the Ly\textalpha Forest]{A New Precision Measurement of the Small-Scale Line-of-Sight Power Spectrum of the L\MakeLowercase{y\textalpha} Forest}

\author{Michael Walther\altaffilmark{1,2,3},
\and Joseph F. Hennawi\altaffilmark{1,2},
\and Hector Hiss\altaffilmark{2,3},
\and Jose O\~norbe\altaffilmark{2},
\and Khee-Gan Lee\altaffilmark{4},
\and Alberto Rorai\altaffilmark{5},
\and John O'Meara\altaffilmark{6}
}
\email{mwalther@physics.ucsb.edu}

\altaffiltext{1}{Physics Department, Broida Hall, University of California Santa Barbara, Santa Barbara, CA
93106-9530, USA}
\altaffiltext{2}{Max-Planck-Institut für Astronomie, Königstuhl 17,
69117 Heidelberg, Germany}
\altaffiltext{3}{International Max Planck Research School for Astronomy \&
Cosmic Physics at the University of Heidelberg}
\altaffiltext{4}{Lawrence Berkeley National Laboratory, 1 Cyclotron Rd, Berkeley, CA 94720, USA}
\altaffiltext{5}{Cambridge Institute of Astronomy, Madingley Rd, Cambridge CB3 0HA, UK}
\altaffiltext{6}{Saint Michael's College, One Winooski Park, Colchester, VT 05439, USA}

\begin{abstract}
We present a new measurement of the Ly$\alpha$ forest power spectrum at
$1.8<z<3.4$ using 74 Keck/HIRES and VLT/UVES high-resolution,
high-S/N quasar spectra.  We developed a custom pipeline to measure
the power spectrum and its uncertainty, which fully accounts for
finite resolution and noise, and corrects for the bias induced by
masking missing data, DLAs, and metal absorption lines. Our
measurement results in unprecedented precision on the small-scale
modes $k>\SI{0.02}{\skm}$, unaccessible to previous SDSS/BOSS analyses.  It is
well known that these high-$k$ modes are highly sensitive to the
thermal state of the intergalactic medium, however contamination by narrow metal lines
is a significant concern.  We quantify the effect of metals on the
small-scale power, and find a modest effect on modes with $k<\SI{0.1}{\skm}
$. As a result, by masking metals and restricting to $k<\SI{0.1}{\skm}$ their impact is completely mitigated.  We present an end-to-end
Bayesian forward modeling framework whereby mock spectra with the same
noise, resolution, and masking as our data are generated from
Ly$\alpha$ forest simulations.  These mocks are used to build a custom
emulator, enabling us to interpolate between a sparse grid of models
and perform MCMC fits.  Our results agree well with BOSS on scales $k<\SI{0.02}{\skm}$ where the measurements overlap. The combination of BOSS' percent level
low-$k$ precision with our $5-15\%$ high-$k$ measurements, results in
a powerful new dataset for precisely constraining the thermal history
of the intergalactic medium, cosmological parameters, and the nature of dark
matter. The power spectra and their covariance matrices are
provided as electronic tables.
\end{abstract}

\section{Introduction}

The \ac{Lya} forest \citep{gunn1965DensityNeutralHydrogen,Lynds1971Absorption-LineSpectrum} is the premier probe of diffuse baryons in the \ac{IGM} at high redshifts.
Its fluctuations can be accurately described in the current \(\Lambda \mathrm{CDM}\) framework and while on large scales it is mostly sensitive to cosmology and structure formation, on small scales it probes the thermal state of the \ac{IGM} due to two effects: Doppler broadening due to thermal motions and pressure smoothing (sometimes called "Jeans" broadening), which affects the underline baryon distribution and depends on the integrated thermal history of the \ac{IGM} \citep{gnedin1998ProbingUniverseLyalpha, kulkarni2015CharacterizingPressureSmoothing,Onorbe2017Self-consistentModeling}.

The major process heating the \ac{IGM} is considered to be reionization.
In this evolutionary phase neutral hydrogen and helium are ionized thus finally allowing photons to freely travel through the Universe.
Excess energy from ionizing photons is distributed into the intergalactic gas leading to strong thermal evolution during reionization.
Studying the \ac{IGM} during reionization provides important insight into the physical state of baryons at that time and thus allows to constrain the environment in which galaxies and \acp{QSO} form.

Hydrogen reionization (as well as \HeI{} reionization which has a similar ionization threshold) was mostly complete by redshift \(z=6\) \citep{fan2006Constrainingevolutionionizing,becker2015Evidencepatchyhydrogen} with a likely midpoint around \(8.5\) \citep{PlanckCollaboration2016Planckintermediate}.
Although it is widely accepted that ionizing photons from stars in
the first galaxies are responsible for reionizing \ion{H}{1} \citep[e.g.][]{Bouwens2015UVLUMINOSITY,Oesch2014MOSTLUMINOUS},
a competing scenario whereby reionization
was driven by faint AGN has also recently been debated (\citealt{madau2015CosmicReionizationPlanck} {, \citealt{Khaire2016RedshiftEvolution}}, but see \citealt{DAloisio2017contributionactive}) as larger samples of quasar candidates were obtained (but not always spectroscopically confirmed) at possibly high redshifts (\citealt{Giallongo2015FaintAGNs} vs. \citealt{Weigel2015systematicsearch}).

For \HeII{}, however, reionization is delayed until \(z\sim 4\) because of its much higher ionization threshold of \(\SI{54.4}{\electronvolt}\).
As stars typically not producing enough photons above this energy \citep{Stanway2016Stellarpopulation,Topping2015EFFICIENCYSTELLAR} quasars are assumed to be the major source of \HeII{} reionization, delaying the full ionization of helium  until quasars become sufficiently abundant \citep{Furlanetto2008historymorphology,mcquinn2009HeIIReionization}.
While observations of the \HeII{} Ly\(\alpha\) forest show that \HeII{} has to be ionized at \(z=2.7\) \citep{worseck2011EndHeliumReionization} with a large fraction already ionized at \(z=3.4\) \citep{Worseck2016EarlyExtended}
there are only indirect constraints on the start of this process by analyses of quasar proximity zones \citep{Bolton2010firstdirect,bolton2012Improvedmeasurementsintergalactic}. Due to the
observational challenges of working in the \ac{UV}, as well as the
universe being mostly opaque to these photons at \(z > 4\), direct
observations of \ion{He}{2} at earlier times will have to wait for the
next generation of space telescopes.
However, the signatures of reionization imprinted on the thermal state of the \ac{IGM} are still observable.

In the standard picture of thermal evolution cold \ac{IGM} gas (few \(\si{\kelvin}\)) is strongly heated during \HI{} and \HeI{} reionization (by few times \(\SI{1e4}{\kelvin}\)), subsequently cools and then experiences additional heating during \HeII{} reionization \citep{mcquinn2009HeIIReionization,compostella2013imprintinhomogeneousHe,puchwein2015photoheatingintergalacticmedium,uptonsanderbeck2016Modelsthermalevolution,McQuinn2016intergalactictemperature-density,Onorbe2017Self-consistentModeling}.
The combined effects of photoionization heating,  Compton cooling, and adiabatic cooling due to the expansion of the universe lead to a net cooling of intergalactic gas between and after the reionization phases which has so far not been conclusively observed.
The combination of those effects also results in a tight power law temperature-density relation for most of the \ac{IGM} gas \citep{hui1997Equationstatephotoionized,puchwein2015photoheatingintergalacticmedium,McQuinn2016intergalactictemperature-density} long after reionization events:
\begin{equation}
\label{eq:t-rho-relation}
 T(\Delta) = T_0 \Delta^{\gamma-1}
\end{equation}
for overdensity \(\Delta = \rho / \bar{\rho}\), temperature at mean density \(T_0\) and an expected slope \(\gamma\approx 1.6\).
During reionization events this slope is expected to be shallower due to the additional
photoionization heating which may also result in a large scatter and a more complicated
density dependence \citep{compostella2013imprintinhomogeneousHe, McQuinn2016intergalactictemperature-density}.
This would be mostly caused by reionization not occurring uniformly, but being intrinsically patchy and therefore leading to significant fluctuations in the \ac{UVB} \citep{Davies2016Largefluctuations,Suarez2017Large-scalefluctuations} as well as temperatures \citep{DAloisio2015LARGEOPACITY} during reionization events.
Hydrodynamical simulations show that within several hundred \(\si{\mega\year}\) the gas relaxes to the tight power law relation of \autoref{eq:t-rho-relation}.
Unfortunately there is so far no consensus about the thermal evolution of the \ac{IGM} during and after \HeII{} reionization, although many measurements have been performed.

Several statistical properties of the \ac{Lya} forest can be used to characterize the thermal state of the \ac{IGM} typically by constraining the smoothness of the forest.
Measurements have been performed using the flux \ac{PDF} \citep{bolton2008Possibleevidenceinverted,Viel2009Cosmologicalastrophysical,Lee2015IGMConstraints,Rorai2017Exploringthermal}, wavelet decompositions of the forest \citep{theuns2002Temperaturefluctuationsintergalactic,lidz2010MeasurementSmallscale,garzilli2012intergalacticmediumthermal}, the curvature of the smoothed transmission \citep{becker2011DetectionextendedHe,boera2014thermalhistoryintergalactic}, the quasar pair phase angle distribution \citep{rorai2013NewMethodDirectly,Rorai2017Measurementsmall-scale}, and decomposition of the forest into individual absorption lines \citep{haehnelt1998Probingthermalhistory,schaye2000thermalhistoryintergalactic,bryan2000DistributionLyForest,ricotti2000EvolutionEffectiveEquation,mcdonald2001MeasurementTemperatureDensity,rudie2012TemperatureDensityRelation,bolton2014consistentdeterminationtemperature}.
There is a new measurement of thermal parameters using the last approach \citep{hiss2016newmeasurementthermal} on the same dataset we'll use in this work that also highlights some systematics of this technique (e.g. due to blending of absorption lines).

In the past the different types of thermal state measurements led to very different results with temperatures at mean density varying by a factor of \(\sim 2\) and \(\gamma\) ranging from \(\gamma<1\) (i.e. an inverted \(T-\rho\) relation) in several PDF analyses \citep{bolton2008Possibleevidenceinverted,Viel2009Cosmologicalastrophysical} that might need new physics to be explained (\citealt{puchwein2012Lymanforestblazar}, but also see \citealt{Rorai2017Exploringthermal} for an alternative explanation based on different sensitivities of the techniques) to the expected asymptotic value.
On the other hand, the curvature technique provides the so far strongest temperature constraint at some characteristic overdensities \(\Delta_\star\), but is insensitive to \(\gamma\) \citep[but see][for an updated approach]{boera2016Constrainingtemperaturedensity} and one therefore needs to rely on external measurements of \(\gamma\) to assess \(T_0\).

To overcome those issues we measure the small-scale (high wavenumber \(k\)) cutoff of the \ac{Lya} forest flux power spectrum to determine the thermal state of the \ac{IGM}.
On smaller scales than this cutoff there is no structure left in the \ac{Lya} forest due to a degenerate combination of spectral resolution as well as thermal and cosmological properties of the \ac{IGM}.
But compared to the other methods this is not only a smoothness measurement (as there are constraints on large scales as well) which allows for breaking of degeneracies.
While the thermal history of the IGM has the strongest effect on the cutoff scale the small scale power is also sensitive to the nature of dark matter \citep{viel2013Warmdarkmatter,Irsic2017Newconstraints} and the power spectrum in general can be used to deduce constraints for a variety of cosmological parameters, e.g. the mass of neutrinos \citep{Palanque-Delabrouille2015Neutrinomasses,yeche2017Constraintsneutrinomasses,baur2017ConstraintsLybackslash}.

The existing measurements of the  \ac{Lya} forest power spectrum  can be divided into two groups.
Either they were obtained from small datasets of high-resolution spectra \citep{mcdonald2000ObservedProbabilityDistribution,croft2002PreciseMeasurementMatter,kim2004powerspectrumflux,viel2008HowColdCold,viel2013Warmdarkmatter} and therefore have sparse redshift sampling as well as lacking precision (especially for large scale modes). Or they have been determined using large datasets from  the \ac{SDSS} \citep{mcdonald2006LyForestPower} or \ac{BOSS} surveys \citep[][, hereafter \citetalias{palanque-delabrouille2013onedimensionalLy}]{palanque-delabrouille2013onedimensionalLy}, but lack the spectroscopic resolution needed to measure the small-scale cutoff of the \ac{IGM} flux power spectrum.
While the latter lead to high precision (\(\sim 2\%\)) measurements they only weakly constrain the thermal state of the IGM on their own due to the limitation in spectral resolution.
On the other hand previous high-resolution measurements have been used to put constraints on the thermal state of the IGM \citep{zaldarriaga2001ConstraintsLyForest}, but typically lack the precision to constrain cosmological parameters.
There have also been recent measurements using medium resolution X-SHOOTER data \citep{irsic2016Lymanalphaforestpower} at \(3<z<4.2\) that in principle probe the \ac{IGM} small scale power.
Lastly, there have been HST/\ac{COS} observations to probe the low-redshift IGM \citep{gaikwad2016IntergalacticLymancontinuum} which is not accessible from the ground due to the atmospheric UV cutoff.

In this work we perform a new power spectrum analysis on a large sample of archival high-resolution spectra and combine this with the existing low- and medium-resolution measurements to enable an accurate measurement of thermal evolution in the \ac{IGM} in a follow-up study \citep{waltherinprepConstraintsthermalevolution}.
This paper is organized as follows.  The dataset of high-resolution
spectra is described in \autoref{sec:dataset}. There we also
discuss our additional treatment of the data to remove metal line contaminants.
In \autoref{sec:ps-measurement-method} we present our approach to measure the power spectrum and describe our forward modeling procedure required to accurately interpret
our measurement. In  \autoref{sec:results} we present our new power spectrum measurement,
quantify the impact of metal line contamination, and compare our results to previous
work. Our conclusions and outlook are given in \autoref{sec:conclusions}.

\section{High-Resolution Quasar Dataset} \label{sec:dataset}

In this section we will describe the dataset we used for our measurement.
First we explain how our quasar sample was constructed.
Then we describe which parts of the selected data were used and how we masked regions of the data to remove contaminants like e.g. metals or \acp{DLA}
Finally we will explain how we regulate the mean flux of our spectra.

\subsection{Dataset}

\begin{deluxetable}{lcc}
\tabletypesize{\footnotesize}
\tablecolumns{3}
\tablewidth{0pt}
\tablecaption{UVES spectra from \citet{dallaglio2008unbiasedmeasurementUV} used for our analysis}
\tablehead{
\vspace{-0.15cm}&&\colhead{median}\\
\colhead{Object}\vspace{-0.15cm}&\colhead{\(z_{\mathrm{QSO}}\)}&\\
&&\colhead{S/N per \(\SI{6}{\kms}\)}
}
\startdata
 \label{tab:Aldo}
 \input{uves_table}
\enddata
\end{deluxetable}

\begin{deluxetable}{lcc}
\tabletypesize{\footnotesize}
\tablecolumns{3}
\tablewidth{0pt}
\tablecaption{HIRES spectra from KODIAQ \citep{omeara2015FirstDataReleasea} used for our analysis}
\tablehead{
\vspace{-0.15cm}&&\colhead{median}\\
\colhead{Object}\vspace{-0.15cm}&\colhead{\(z_{\mathrm{QSO}}\)}&\\
&&\colhead{S/N per \(\SI{6}{\kms}\)}
}
\startdata
 \label{tab:OMeara}
 \input{kodiaq_table}
\enddata
\tablenotetext{a}{objects are part of DR2, but a pre-DR2 reduction has been used}
\tablenotetext{b}{objects are not part of DR1 or DR2, but reduced in the same way}
\tablenotetext{c}{objects are part of DR1, but a pre-DR1 reduction has been used}

\end{deluxetable}

Our measurement of the power spectrum was performed using 38 high-resolution quasar spectra (see \autoref{tab:Aldo}) from \citet{dallaglio2008unbiasedmeasurementUV} observed with the \acl{UVES} (UVES, \citealt{dekker2000Designconstructionperformance}) at the \ac{VLT}, and 36 spectra (see \autoref{tab:OMeara}) from the \ac{KODIAQ} project \citep{lehner2014GALACTICCIRCUMGALACTICVI} observed with the \acl{HIRES} (HIRES, \citealt{vogt1994HIREShighresolutionechelle})  at Keck.
For the latter we used the highest S/N part of DR1 \citep{omeara2015FirstDataReleasea} and additional data beyond DR1 \citep[mostly early reductions of objects in DR2][]{omeara2017kodiaqdr2} reduced in the same way. {Reduced and continuum fitted spectra of all UVES and KODIAQ DR1 data used here are available in the \texttt{igmspec} package \citep{Prochaska2017igmspec}, KODIAQ DR2 data will be available in future \texttt{igmspec} releases and on \href{https://koa.ipac.caltech.edu/applications/KODIAQ/}{the KODIAQ database webpage}\footnote{\href{https://koa.ipac.caltech.edu/applications/KODIAQ/}{https://koa.ipac.caltech.edu/applications/KODIAQ/}}.}.

Most of the UVES spectra have full coverage between the atmospheric cutoff at \(\lambda_\mathrm{obs}\sim\SI{3100}{\angstrom}\) and \(\lambda_\mathrm{obs}\sim\SI{1}{\micro\meter}\).
This allows us to use a large range of spectrum redward of the Ly$\alpha$ forest to search for metal lines as well as enabling us to search for \acp{LLS} using higher order Lyman series transitions, in many cases even exploiting coverage of the Lyman-limit.

For the \ac{KODIAQ} data the typical red spectral coverage ends at \(\lambda_\mathrm{obs}\sim\SI{6000}{\angstrom}\), while the blue spectral cutoff is comparable to UVES \(\lambda_\mathrm{obs}\sim\SI{3100}{\angstrom}\).
For a few cases in both datasets, however, even the \ac{Lya} forest was not fully covered.

\begin{figure}
\centering
\plotone{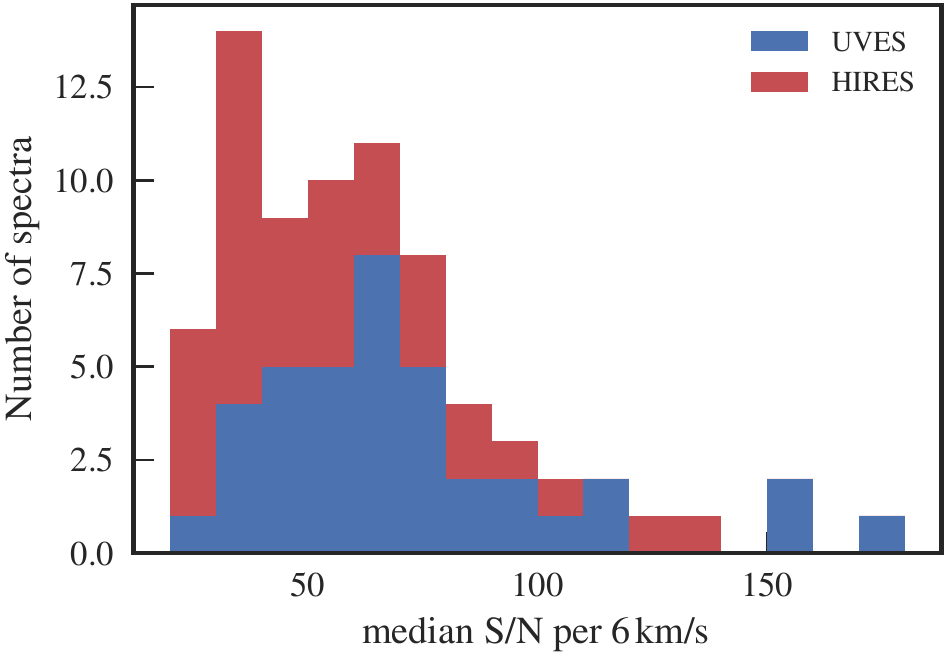}
\caption{Histogram of the median S/N per \(\SI{6}{\kms}\) in the Ly\(\alpha\) forest region used divided by dataset. Note that we only used spectra with \(\mathrm{S/N}>20\) in the dataset}\label{fig:redshift-sn-dist}
\end{figure}

The objects used in our analysis were chosen to have a median \(\textup{S/N}>20\) per \(\SI{6}{\kms}\) interval inside the Ly$\alpha$ forest region covered by the spectra.
We also chose to omit spectra with known \acp{BAL}.
Finally we omitted sightlines with \(3.0<z<3.5\) that were color selected to avoid potential biases due to their increased abundance of \acp{LLS} \citep{worseck2011GalexFarUltraviolet}.

The distribution of S/N for the dataset used in our analysis is shown in \autoref{fig:redshift-sn-dist}.
Many objects have a much larger S/N than our cut (up to about a median S/N of \(150\textup{ per }\SI{6}{\kms}\) inside the \ac{Lya} forest in a sightline).
This very high S/N enables us to perform the measurement without strong systematics due to the limited accuracy of our noise model affecting our measurement at the scales of interest.

The nominal FWHM spectral resolution of the data varies between \(\SI{3.1}{\kms}\) and \(\SI{6.3}{\kms}\) with a typical value around \(\SI{6}{\kms}\).
Therefore all \ac{Lya} absorption lines are resolved and we expect thermal broadening and broadening due to pressure smoothing to be the effects determining the smoothness of lines, and not smoothing due to finite spectroscopic resolution \citep[see e.g.][for a measurement of line width for thermally broadened lines]{bolton2014consistentdeterminationtemperature}.

The data was already reduced and continuum fit, and we briefly summarize the details here.
The UVES data were fit with both a global power law in non-absorbed regions and local cubic splines fitted automatically with spline point separations depending on continuum slope.
Systematic biases in this technique were estimated and corrected using a Monte Carlo analysis on mock data
by \citet{dallaglio2008unbiasedmeasurementUV}.
The \ac{HIRES} continua were hand-fitted by John O'Meara one echelle order at a time by placing Legendre polynomial anchor points at non-absorbed positions \citep[for estimates of continuum fitting errors, see][]{kirkman2005HIopacityintergalactic,faucher-giguere2008DirectPrecisionMeasurement}.
Afterwards a 4th to 12th order polynomial was fit through the anchors.
Further details about the reduction and continuum fitting techniques can be found in the respective data papers \citep{dallaglio2008unbiasedmeasurementUV,omeara2015FirstDataReleasea}.

We will use this high-resolution dataset to measure the small scale power spectrum of the \ac{Lya} forest in redshift bins of size \(\Delta z=0.2\) with central redshifts between \(\bar{z}=1.8\) and \(\bar{z}=3.4\).
Each of the bins contains at least eight quasar spectra, with the majority (all but the low and high redshift edges of our sample) containing more than 14 quasar spectra.
The datasets also contain eight spectra that cover higher redshifts which we did not analyze due to the small amount of data available (two spectra cover \(z\gtrsim 4.0\), the rest only cover the forest for \(z<3.6\) just half a bin further than our analysis).

In this work and our companion paper \citep{waltherinprepConstraintsthermalevolution} we compare our power spectrum measurements to measurements from
lower spectral resolution data from BOSS \citepalias{palanque-delabrouille2013onedimensionalLy} and from the XQ-100 survey \citet{irsic2016Lymanalphaforestpower} based on X-SHOOTER data, and
also conduct joint model fits. To facilitate this comparison we use the same binning in redshift from \(\bar{z}=2.2\) to \(\bar{z}=3.4\).

\subsection{Spectral Masking Procedure}
\label{sec:dataprep} \label{sec:mask}

To prepare the data for the power spectrum computation we
restrict our attention to the restframe wavelength range of \(\SI{1050}{\angstrom} < \lambda_r < \SI{1180}{\angstrom}\).
This was done to exclude the \ac{Lya} proximity zone, accounting for possibly
large redshift errors, as well as to exclude the Ly\(\beta\) and \ion{O}{6} \(\lambda 1035\) emission lines and possible blueshifted absorption from these as well as increased continuum fitting errors close to emission lines.
This is the same range used in \citetalias{palanque-delabrouille2013onedimensionalLy} and is considered a conservative choice for the \ac{Lya} forest region.

We masked parts of the spectrum to reduce contaminations due to low-quality data, high-column
density absorbers such as \acp{DLA}, and metal lines.
If a pixel is already masked during the reduction (due to e.g. cosmic rays or gaps in spectral coverage) it stays masked.
According to \citet{mcdonald2005PhysicaleffectsLy} excluding \acp{DLA} and \acp{LLS} from
power spectrum calculation only changes the power by \(< 2\%\) on the scales of interest for our analysis.
We nevertheless excluded absorbers with clearly visible damping wings, i.e.
\acp{DLA} and super-LLSs, from our spectra to make sure those do not influence the result.
For this we masked the core and wings of these strong absorbers by eye until the wings
are below the noise level of the spectra.
In most cases this leads to the exclusion of big continuous spectral regions at the boundary of a redshift bin or the whole data of this object falling inside a redshift bin.
Where a \ac{DLA} mask removed only the bin center we used the longer of the two remaining spectral regions at the bin boundaries to compute the power.
Therefore spectra with \acp{DLA} are only shorter, but with no additional gaps in the data.
Finally we removed spectra that were shorter than 10\% of a redshift bin corresponding to a minimum length of \(\Delta z\sim0.02\) (or \(\SI{3500}{\kms}\)) to avoid noisy contributions from short spectra that consist of only few absorption lines.

\begin{deluxetable}{rSrS}
\tabletypesize{\footnotesize}
\tablecolumns{2}
\tablewidth{0pt}
\tablecaption{List of metal transitions that were masked}
\tablehead{\colhead{Ion}&\colhead{\(\lambda_\mathrm{rest}/\si{\angstrom}\)}&\colhead{Metal transition}&\colhead{\(\lambda_\mathrm{rest}/\si{\angstrom}\)}}
\startdata
\label{tab:absorbers}
\input{metallist}
\enddata
\tablenotetext{a}{strongest transitions, therefore used for all the masking techniques}
\end{deluxetable}

Metal lines in the \ac{Lya} forest are expected to increase the power primarily on small scales  \citep[\(k\gtrsim \SI{0.1}{\skm}\), see][]{mcdonald2000ObservedProbabilityDistribution,lidz2010MeasurementSmallscale}
due to their narrower widths compared to \ac{Lya},
but due to correlations induced by the rest-frame velocity separations of different transitions (e.g. between \SiIII{} and \HI-\ac{Lya} or between the \ion{C}{4} doublet lines) there is contamination on larger scales as well \citep{croft1999Powerspectrummass,mcdonald2006LyForestPower}.

To reduce the impact of metal absorption inside our sightlines we masked metal lines in the forest region.
We first identified metal absorption lines in the Ly\(\alpha\) forest by having two of the authors (H. Hiss and M. Walther) visually inspect the spectra and mask metal contamination in several ways.
First, we look for \acp{DLA} inside the forest and mask all strong metal absorption lines corresponding to the \ac{DLA} redshifts.
For this we used all metal transitions in \autoref{tab:absorbers} and masked a region of \(\SI{60}{\kms}\) in each direction around each identified absorber.
Then we search for absorption from common doublet transitions (\ion{Si}{4}, \ion{C}{4}, \ion{Mg}{2}, \ion{Al}{3}, \ion{Fe}{2}) redwards of the forest where the spectrum is mostly clean and mask all associated metal lines analogous to our procedure for \acp{DLA} (using the same metal catalog and mask width), until there are no doublet features left redward of the forest.
We also masked out a region of \(\SI{200}{\kms}\) in each direction around redshift zero to get rid of metal contamination from the Milky Way (in this case the only relevant transition is \ion{Ca}{2} for \(2.33<z_\mathrm{QSO}<2.78\)).

Additionally, we used an automated \ac{pLLS} finder written by John O'Meara to identify strong absorption systems.
This finder works by searching for pixels with zero flux (within some threshold) at the corresponding positions of \ac{Lya}, \textbeta, \textgamma{} and higher Lyman transitions if available and grouping them into systems of \ac{LLS} candidates.
The candidates were then visually inspected by one of the authors (John O'Meara) and compared to theoretical line profiles of absorbers with \(\log(N_\mathHI)=15,16,17\) in \ac{Lya} to Ly\textgamma, and systems which appeared consistent were positively identified
as pLLSs.
For these systems, associated metal absorbers from a reduced line list (see \autoref{tab:absorbers} absorbers marked with a) were masked with the same velocity window size as
above.
Note that the hydrogen absorption arising from the pLLSs  identified in this way
was not masked regardless of their \(N_\textrm{HI}\).

Lastly we perform a line-fitting analysis \citep[see][]{hiss2016newmeasurementthermal} using a semi-automatic wrapper around \texttt{VPFIT} \citep{carswell2014VPFITVoigtprofile} on the same set of \ac{Lya} spectra.
The result of this is a distribution of line widths \(b\) and column density \(N_\mathHI\) for all fitted lines assuming that all absorption is due to hydrogen.
For \HI{} gas at a given column density lines are broadened both thermally (due to finite pressure broadening and instantaneous temperature Doppler broadening) and hydrodynamically due to local gas motions.
There is a minimum broadening \(b_{\rm cut}(N_\mathHI)\) populated by absorbers with
zero line-of-sight peculiar velocity which are purely thermally broadened
\citep[see e.g.][for more details]{schaye2000thermalhistoryintergalactic,hiss2016newmeasurementthermal}.
Therefore \(b(N_\mathHI)\) should cut off for \(b<b_{\rm cut}(N_\mathHI)\) and all remaining lines narrower than this cutoff can be attributed to fitting artifacts at the edges of strong absorbers, noise fluctuations, or narrow metal absorption lines.
This cutoff is fit using an iterative procedure similar to the one used in \citet{rudie2012TemperatureDensityRelation}.
We identified all the lines in previously unmasked spectral regions fulfilling \(b<\SI{11}{\kms} (N_\mathHI / 10^{12.95}\si{\per\centi\meter\squared})^{0.15}\) as these are narrower than the thermal cutoff \citep[see][for details]{hiss2016newmeasurementthermal}.
For each of these lines (that are in the \ac{Lya} forest region) we checked if they could be identified using a second metal transition clearly lining up at the same redshift. In practice, these identifications were most easily made when both metal transitions form a doublet.
Given a positive identification, masking was then performed as for metal absorbers that we identified redward of the \ac{Lya} forest, i.e. determining the redshift of the absorber and using the full list of metal lines in \autoref{tab:absorbers}.

Neither of these techniques produces a fully metal free \ac{Lya} forest, and we briefly elaborate on these limitations.
The search for metals redwards of the forest only finds transitions if both doublet counterparts are visible at redder wavelength than the \ac{Lya} emission line.
If the doublet counterpart of a line falls into a spectral region that was not covered or is blended with a different line (from either a different
absorber redshift or e.g. a telluric absorption feature) the contaminating system will often go undetected.
For example, for a \ion{C}{4}\(\lambda 1548/1550\)
system at \(z>2.5\) that might contaminate the forest, the \ion{Mg}{2} \(\lambda 2796/2803\) doublet would land at \(\lambda_\mathrm{obs} > \SI{1}{\micro\meter}\) and hence outside of the spectral coverage of our spectra, so only if the absorber shows \ion{Fe}{2} or \ion{Al}{3} doublets would this \ion{C}{4} absorber be masked.
On the other hand especially for the higher redshift bins a large fraction of the spectral range redwards of the forest is contaminated by telluric absorption making doublet identification very challenging.
The largest problem for this method is therefore limited usable spectral coverage in the red.

For the automated \ac{pLLS} finder at least \ac{Lya} to Ly\textgamma{} need to be detectable to identify a \ac{pLLS}.
This leads to a minimal redshift of \(z\gtrsim2.1\) for absorbers that can be identified as they need to be at observed wavelength higher than the atmospheric cutoff \(\lambda_\mathrm{obs}\approx \SI{3000}{\angstrom}\).
For reduced spectral coverage in the blue this minimal redshift is correspondingly higher.

The last of our metal masking procedures only recognizes metal absorption lines significantly narrower than  the cutoff in the \(b(N_\mathHI)\) distribution and requires a second metal transition for identification.
Therefore singlet lines are not masked by this technique unless another transition from a different metal species clearly lines up with them.
Also metal absorbers were not removed if all components are broadened above the \(b(N_\mathHI)\) threshold adopted, and are therefore not recognized as metal candidates.

\begin{figure*}
\plotone{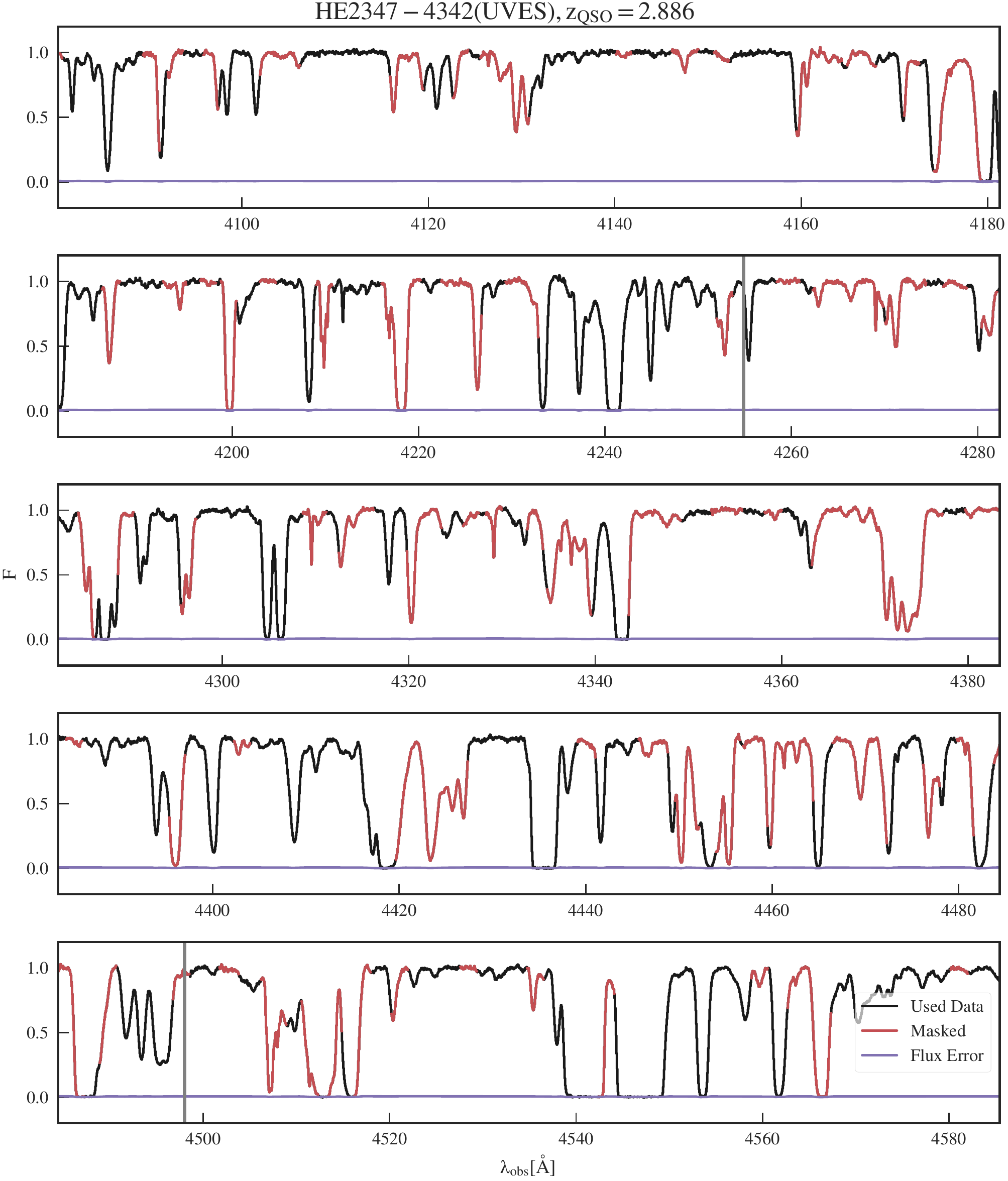}
\caption{The \ac{Lya} forest region for one of our spectra (HE2347-4342) with regions masked due to possible metal contamination in red.
The purple line shows the error level of the spectrum.
Gray vertical lines show the boundaries of our redshift bins.
Note that due to our approach not only metals are masked, but also coincidental pieces of the \ac{Lya} forest.}\label{fig:example-spec}

\end{figure*}

\begin{figure*}
\centering
\plotone{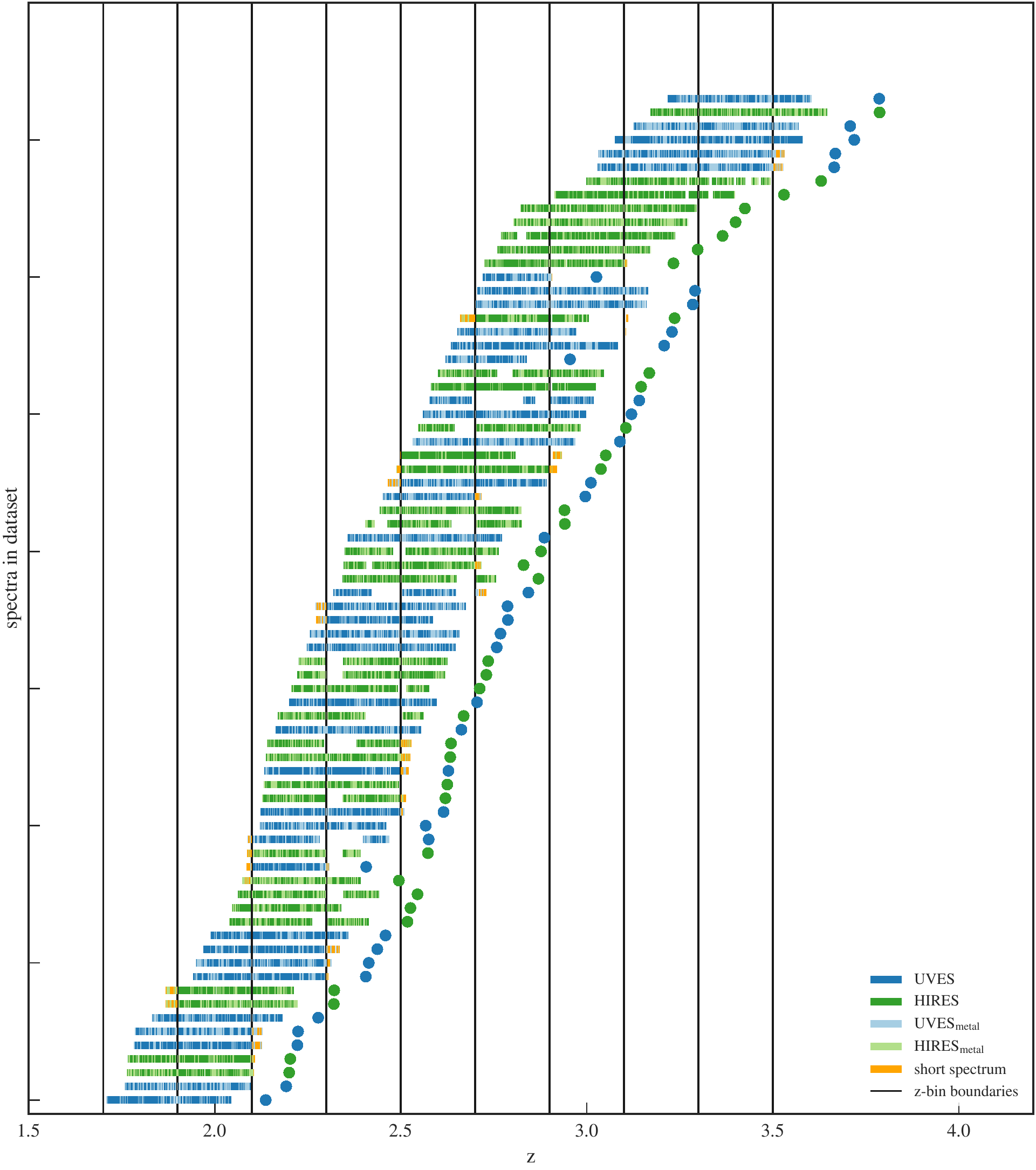}

\caption{
Redshift coverage of the dataset used colored by spectrograph.
Blue (green) lines show the spectral coverage of each used spectrum in the \ac{Lya} forest for the UVES (HIRES) subset.
Circles mark the corresponding quasar redshifts.
Most of the long gaps in this figure are due to missing data between e.g. non-overlapping echelle orders or masked out \acp{DLA} while the light colored parts show masks due to possible metal contamination (see \autoref{sec:mask}). Orange lines show regions that were ultimately rejected because of limited forest coverage. Vertical lines mark boundaries of the redshift bins used in the analysis.}\label{fig:redshift-lines}

\end{figure*}

\begin{figure}
\plotone{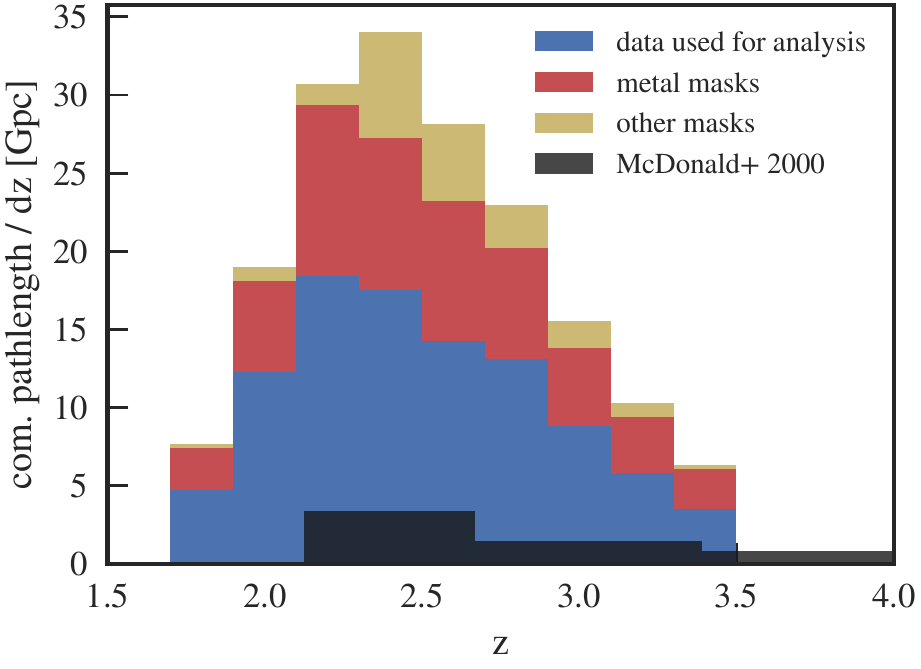}
\caption{The pathlength of used data in our analysis compared to the pathlength of spectra masked due to several reasons (metals: red, other: yellow). About 30\%-40\% of each redshift bin were masked due to possible metal contamination. Far less due to other things, e.g. \acp{DLA}, bad reductions or only short available pathlength.
As a comparison the amount of data in  \citet{mcdonald2000ObservedProbabilityDistribution} is shown in black for their redshift binning.
Normalization is such that the area (and not the height) of each bar corresponds to the total path inside it.}\label{fig:metal-mask-lines}

\end{figure}

In \autoref{fig:example-spec} we show an example to illustrate how much of a typical spectrum is masked.
The black line shows the non-masked part of the spectrum while the red line shows parts that are masked due to possible metal contamination.  One can see that many narrow lines in the spectrum are masked, but also that many regions of the forest coincidentally overlap with positions where metal lines associated with our identified absorbers could lie, but that don't actually contain any visible metal absorption.

In \autoref{fig:redshift-lines} we present an overview of the dataset.
For each quasar spectrum the emission redshift is shown as well as the full coverage of the spectrum after masking pre-existing gaps in the data and \acp{DLA} (which therefore appear as white gaps).
The remaining spectrum is divided into the data used (dark colors), masked data due to possible metal contamination (bright colors), and spectra not used due to failing our requirement of spectral extent (yellow).
We also show the usable \ac{Lya} forest pathlength after applying the full masking procedure compared to the available pathlength when not masking metals in \autoref{fig:metal-mask-lines}.
Up to \(\sim 40\% \) of the forest gets removed by applying our metal masking procedure strongly reducing our dataset size and therefore the precision of our measurement results.
Although not perfect, our procedure significantly reduces the metal line contamination in our spectra, which decreases the amount of small-scale power compared with an unmasked dataset as we will see in \autoref{sec:metalcomp}.
However, this masking also changes the power spectrum due to the application of a complex window function in configuration space.
To correct for this effect we forward model the masking (see \autoref{sec:fwmodel}).

\subsection{Mean Flux Regulation and Continuum Uncertainties}
\label{sec:mfreg}

In principle the estimation of quasar continua in the data is subject to errors as well.
We perform our power spectrum measurement on the flux contrast
\begin{equation}
\delta_F=\frac{F-\bar{F}}{\bar{F}}.
\label{eq:fluxcontrast}
\end{equation},
with \(\bar{F}\) being the mean transmission of the \ac{Lya} forest
Because of this, any global misplacement of the continuum will be divided out as long as the mean transmission measurement we divide by is measured on the same spectrum.
In addition, incorrect placement of the continuum could lead to gradients or wiggles
in the data that could source additional large scale power
(on scales \(k<\SI{1e-3}{\skm}\)). However,  this will not strongly impact the small scale power measurement we want to perform in this work\footnote{We performed tests on model simulations at \(z=3\) including quasar continua. Comparing the power spectrum using the true continuum vs. a hand-fitted continuum showed that only the very largest scales (smallest k) were affected by the procedure.}\citep{lee2012Meanfluxregulated}.

Nevertheless, we perform a mean flux regulation on the dataset using the technique  of \citet{lee2012Meanfluxregulated} which enables us to easily divide out the mean transmission as well as possible gradients in the continuum fits.
For this the transmission of the \ac{Lya} forest region (excluding masked pixels as well as possible proximity regions as discussed earlier) of each quasar sightline is first fit by a linear relation \(f(z)\) times the mean transmission function \(\bar{F}=\exp(-\tau_\mathrm{eff})\) with
\begin{equation}
\tau_\mathrm{eff}=C+\tau_0\left(\frac{1+z}{1+z_0}\right)^\beta \label{eq:mfevo-tau}
\end{equation}
following the functional form and parameters from \citet{becker2013refinedmeasurementmean}.
Afterwards \(f(z)\) is divided out so that the mean flux evolution of each spectrum follows
the same relationship.

The flux contrast \(\delta_F\) is now easily obtained by using \autoref{eq:fluxcontrast} on mean flux regulated spectra using \autoref{eq:mfevo-tau} for the mean flux evolution instead of dividing each spectrum by a mean transmission estimate for this spectrum.
This allows us to divide out the mean flux at each pixel analytically based on the fit of the mean flux evolution across each spectrum.
While this in principle leads to a reduced large-scale power \citep{lee2012Meanfluxregulated}, we are not measuring the affected scales in this work.

The resulting spectra are finally divided into our redshift bins of \(\Delta  z_{\textup{chunk}}=0.2\) (with the first bin starting at \(z=1.7\)) to increase redshift resolution to a level where we could closely monitor thermal evolution.
We will henceforth call these pieces of spectra "spectral chunks".

\section{Power Spectrum Measurement and Forward Modeling Procedure}
\label{sec:ps-measurement-method}
In this section we describe our procedure for measuring the flux power
spectrum.  First we explain how we measure the raw power spectrum. Next we discuss the impact of
our masking (especially the masking of metal contaminants) on this
measurement. After this we discuss the models we use in most of this
work, how we generate mock spectra from them, and approximate the data
covariance matrix by combining information from both data and
models. Finally, we discuss how we create a fast emulator of our model
power spectra, and use this to fit models to our data, allowing us to
correct for the masking.

\subsection{Measuring the Power Spectrum}
\label{sec:ps-definition}

We calculate the power spectrum from the flux contrast \(\delta_F\) defined in \autoref{eq:fluxcontrast}.
As our spectra are not periodic and are not regularly sampled because of masking, we Fourier transform \(\delta_F\) using a Lomb-Scargle periodogram \citep{lomb1976Leastsquaresfrequency,scargle1982Studiesastronomicaltime}, allowing us to compute the raw power \(P_{\mathrm{raw}}(k)\) of each spectrum for a linearly spaced set of modes from the fundamental mode given by the length of the spectral chunk to the Nyquist limit.
We subtract the noise power  \(P_{\mathrm{noise}}(k)\) from this raw power, and divide the difference by the window function \(W_R\) resulting from finite spectroscopic resolution \(R\), following the FFT method described in \citetalias{palanque-delabrouille2013onedimensionalLy}:
\begin{equation}
  P_{\mathrm{data}}(k)  =  \left< \frac{ P_{\mathrm{raw}}(k) - P_{\mathrm{noise}}(k) }{W_R^2(k,R,\Delta v)} \right>,
  \label{eq:power_definition}
\end{equation}
with
\begin{equation}
  W_R(k,R,\Delta v) = \exp\left(- \frac{1}{2}(kR)^2\right)  \frac{\sin(k\Delta v /2)}{(k\Delta v/2)},
  \label{eq:res_correction}
\end{equation}
and \(\Delta v\) refers to the pixel scale of the spectra in velocity units.

The noise power \(P_{\mathrm{noise}}\) is measured by creating 100 realizations of Gaussian random noise generated from the \(1\sigma\) error vector of each quasar spectrum.
The resolution assumed for the window function correction is the nominal slit-resolution of the spectrograph which is different from the actual spectral resolution of the data,
which also depends on the seeing of the observations.
For the typical resolution in our dataset (so a resolving power of 50,000 or equivalently \(R=\SI{2.55}{\kms}\)) we get \(W_R^2(\SI{0.1}{\skm})\sim0.94\). Given
this small correction, even $\sim 20\%$ error in our knowledge of the resolution
(due to the unknown seeing), would only lead to a small $\lesssim 4\%$
correction of the power at \(k<\SI{0.1}{\skm}\) (for further discussion of this
point see \autoref{sec:final-measurement} and \autoref{sec:reseffects}).
Therefore, the resolution uncertainty does not significantly affect our measurement of the \(k\)-modes we consider.

\footnote{The averaging \(\langle\dots\rangle\) is performed over the individual periodograms of all spectral chunks inside a redshift bin, and also the average over all modes \(k\) inside logarithmic bins in \(k\), where equal weights are given to each individual mode from any spectrum. As the fundamental mode of a shorter spectrum is at larger \(k\), there are less modes available in a given band power for shorter spectra and they are therefore effectively downweighted by performing the average over modes.}
We chose to use the same logarithmically spaced \(k\)-bins as in \citet{mcdonald2000ObservedProbabilityDistribution} for our analysis. Note that this is different from the linear spacing adopted by \citepalias{palanque-delabrouille2013onedimensionalLy} and \citep{Irsic2017Newconstraints}.

We adopt the same Fourier normalization convention as the BOSS measurement \citepalias{palanque-delabrouille2013onedimensionalLy}, such that the variance in flux contrast is \(\sigma^2_{\delta_F}=\sigma^2_{F}/\bar{F}^2=\int_{-\infty}^{\infty}P(k)\mathrm{d}k/(2\pi)\).
Note that this differs from the conventions used by some older high-resolution measurements (see \autoref{sec:app_conventions}) leading to additional normalization factors needed when comparing to those results.

\subsection{The Window Function Resulting from Masking}
\label{sec:window-function}

As described in \autoref{sec:dataprep} we masked out parts of the data in real space due to metal contamination.
This is a different approach than the one used by \citetalias{palanque-delabrouille2013onedimensionalLy} who estimated the metal power from transitions with \(\lambda>\lambda_{\mathrm{Ly}\alpha,\mathrm{rest}}\) by measuring the power redwards of the forest in lower redshift quasar spectra.
The power measured in this way was then subtracted from the measurement, but this method can never account for transitions with \(\lambda < \lambda_{\mathrm{Ly}\alpha,\mathrm{rest}}\) which always fall into the forest (e.g. the \SiIII \(\lambda1206\) line that leads to the \SiIII\, correlation feature in those measurements, see \citealt{mcdonald2006LyForestPower}).

Masking spectra in configuration space with a window function  \(W_m\) leads to the measured \(P_\mathrm{masked}\) for each spectrum being effectively a convolution of the true power \(P_\mathrm{true}\) with the square of the Fourier transform \({\tilde W_m}\) of this window function:
\begin{equation}
P_\mathrm{masked} =P_\mathrm{true}*{\tilde W_m}^2.
\label{eq:making_conv}
\end{equation}
Determining the true power thus requires deconvolving the window function or adopting a different technique to measure the power taking into account the windowing \citep[like the minimum variance estimator used in][]{mcdonald2006LyForestPower}.
In this work we opt to use a simpler approach by generating forward models of the data with and without the windowing applied to be able to determine the effect this window function has on our data.
In the end, we correct our data for those effects by dividing the measurement by a correction function based on these models.
In the remainder of this section we describe how our forward models are generated, how we estimate the data covariance matrix based on those models, and how we fit the data using those models.

\subsection{Simulations and Mock Spectra} \label{sec:dm-skewers}

We use simulations of the \ac{Lya} forest for two reasons.
First, we need to simulate the effect of noise, resolution and the window function due to masking on the power spectrum.
Second, we want to connect the information encoded in the power spectrum to a thermal history of the IGM (parametrized by the temperature at mean density \(T_0\), the slope of the temperature-density relation \(\gamma\) and the pressure smoothing scale \(\lambda_P\)).
For this purpose,  one generally needs to run hydrodynamical simulations with different thermal histories.
However, these are computationally expensive and at least for the first point we only need a model that is flexible enough to provide a good
fit the to observed power spectra, but it need not necessarily provide the correct thermal parameters. Because  \ac{DM} only simulations
of the Ly$\alpha$ forest are more flexible and computationally inexpensive to generate, for all the forward modeling in this paper,
we use  approximate \ac{DM} only simulations (a single box with different thermal parameters generated in post-processing) with a semi-numerical approach to paint on the thermal state of the IGM \citep{croft1998RecoveryPowerSpectrum,meiksin2001Particlemeshsimulations,hui1997Equationstatephotoionized,gnedin1998ProbingUniverseLyalpha,gnedin2003LinearGasDynamics,rorai2013NewMethodDirectly}
This fast simulation scheme allows us to generate a grid of \(\sim 500\) combinations
of thermal parameters in a reasonable amount of time.

These DM-based simulations are however, not sufficiently accurate to infer the
thermal state of the \ac{IGM}, as they produce significant biases in thermal
parameters when fitted to mock data based on hydrodynamical
simulations \citep[see e.g.][for a detailed comparison between
  hydrodynamical and dark matter only
  simulations]{sorini2016MODELINGLyFOREST}.
They do, however, provide a good fit to the data (compare lines and same color errorbars in \autoref{fig:dm_data_compare} which we'll discuss in more detail later) and we therefore use them to correct for the window function in our measurement as well as testing our analysis procedure and our fast power spectrum emulator (interpolation scheme, see \autoref{sec:emulator}).
For inference of IGM thermal parameters a grid of hydrodynamical simulations will be used in a companion paper \citep{waltherinprepConstraintsthermalevolution}.

For the \ac{DM} only simulations we use an updated version of the TreePM code described in \citet{white2002MassFunction} to simulate the evolution of dark matter particles from initial conditions at $z=150$ up to $z=1.8$.
We   use a simulation with \(L_\mathrm{box}=\SI{30}{\hinvMpc}\) and \(2048^3\)
particles with a Plummer equivalent smoothing of \(\SI{1.2}{\hinvkpc}\) based on a \citet{planckcollaboration2014Planck2013results} cosmology with \(\Omega_\mathrm{m} = 0.30851, \Omega_\mathrm{b} h^2 = 0.022161, h = 0.6777, n_s = 0.9611 \textrm{ and } \sigma_8 = 0.8288\).
We do not include uncertainties in cosmological parameters in our models as CMB measurement errors on these parameters \citep{hinshaw2013NineYearWilkinson,PlanckCollaboration2016Planck2015} are much smaller than current constraints on thermal parameters.

The model is generated for snapshots with \(z\) between 1.8 and 3.4 with a separation of 0.2, the same as our power spectrum measurements, and provides a dark matter density and velocity field.
However, the relevant quantities for absorption in the IGM are baryonic density and temperature fields.
The results of previous hydrodynamic simulations suggested a computation of the relevant baryonic fields using scaling relations on the dark matter quantities \citep{hui1997Equationstatephotoionized, gnedin1998ProbingUniverseLyalpha, gnedin2003LinearGasDynamics}.
This basically consists of smoothing the \ac{DM} density field with a Gaussian kernel to mimic pressure support as well as rescaling the densities to get the right \(\Omega_b\).
Temperatures are then introduced by applying the power law temperature-density relation from \autoref{eq:t-rho-relation} to the density field \citep[see][section 2.2 for the exact procedure]{rorai2013NewMethodDirectly}.

Given that the UVB is not known perfectly, we created a sequence of models with different mean transmissions \(\bar{F}\) spanning an \(\sim 5 \sigma\) range around the current observational constraints by \citet{becker2013refinedmeasurementmean}, \citet{faucher-giguere2008DirectPrecisionMeasurement} and \citet{kirkman2005HIopacityintergalactic}.
This was done by rescaling the optical depths of the full set of skewers to match the desired \(\bar{F}\).
In total we have a parameter grid of \(\sim500\) different thermal parameter combinations \((T_0,\lambda_P,\gamma)\) with each of those evaluated for 5 different values of \(\bar{F}\)

\subsection{Forward Modeling Approach}
\label{sec:fwmodel}

As the observed spectra are much longer than the simulation box we first divided each spectral chunk of \(\Delta z=0.2\) into regions smaller than our box and assigned a random simulated model skewer to each of the pieces.
The skewers were than assumed to fall on the respective position of the spectrum and truncated to have the same length as the spectral chunk.
A model of a single real spectrum therefore consists of \(4-8\) simulation skewers.
We generated \(\sim 15,000\) skewers of the same length as the simulation box for each parameter combination, therefore we have the equivalent of \(\gtrsim1900\) mock spectra (or equivalently \(\gtrsim100\) times our whole dataset) available at each set of parameters.
This step is required to enable us to add the wavelength dependent noise and masks to the models.

While the overall mean flux of the box is a free parameter of our models, we renormalize the flux in each pixel of the skewer again to account for the slight redshift evolution of the mean flux along the skewer with respect to the mean flux of the simulation snapshot.
To do this the fluxes are converted back to optical depth by \(\tau=-\log(F)\).
We then use the best fit relation to the \(\tau\) evolution by \citet[][see \autoref{eq:mfevo-tau}]{becker2013refinedmeasurementmean} to compute the fractional change in \(\tau_\mathrm{eff}\) between box redshift and the individual pixel redshifts and rescaled \(\tau\) at each pixel with the corresponding value.
After this we convert back to fluxes.
The same mean flux evolution function is then later taken out in the power spectrum computation when we, analogous to our procedure on the data, divide by the mean flux in order to compute the flux contrast.
We then convolved these spectra with the respective instrument resolution, and interpolate them onto the same wavelength grid as the observed data.
  The result of these steps are mock \ac{Lya} forest spectra with the same noise properties, spectral coverage, and masking of our data.
  We henceforth call these the `forward models'.

We also compute the power spectrum of the same number of model skewers without any noise,  masking, or degradation of resolution
which we compare to the power spectrum of our mocks to validate our power spectrum pipeline, and
to determine the window function correction. We will henceforth refer to these
as `perfect models'.

For fitting the BOSS data, there is no need to create full forward models, because all the imperfections our forward model approach treats are already
accounted for by the BOSS power spectrum pipeline, and therefore we simply compare the BOSS data to the perfect models.

\subsection{Covariance Matrix Estimation}\label{sec:covmat}

\begin{figure*}
\centering
\plotone{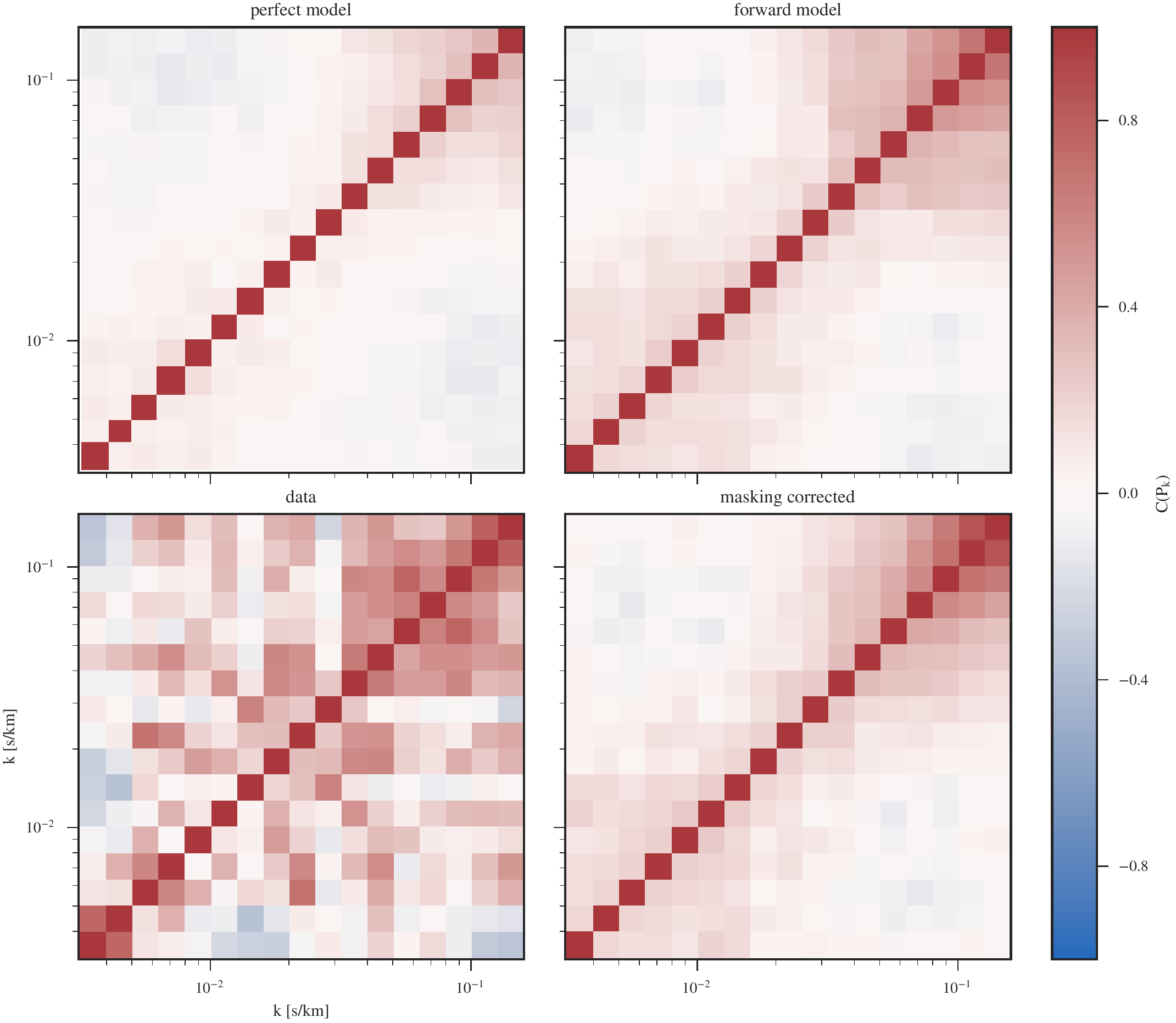}
\caption{The correlation matrices of the non-window corrected, metal-masked \ac{Lya}-forest flux power at \(z=2.8\) as measured from our DM perfect model with parameters close to the best fit (upper left, see \autoref{sec:param_estim} for how the fit was done), the forward model (upper right) at the same parameters, the data (lower left), and the final masking corrected measurement based on the forward model (lower right, see \autoref{sec:raw-measurement} for an explanation about how this was obtained).
We can see that the data correlation matrix is far noisier than the others due to the limited data sample size and therefore not usable for model fitting.
For the forward model correlations we observe that k-bins that are close together are mildly correlated (\(\approx 20\%\)) and bins on very different scales are mildly anticorrelated.
For \(k\gtrsim0.03\) correlations get far stronger due to the power spectrum cutoff as well as the stronger influence of contaminants, e.g. metals, noise and finite spectroscopic resolution.
For the perfect model correlations are far weaker except for neighboring k-bins or bins in the cutoff region.
As the exact position of this increase in correlation depends on the power spectrum cutoff position (and therefore e.g. on thermal parameters) we interpolate between correlation matrices when doing model fitting.
The final masking corrected correlation matrix looks very similar to the forward model case except for additional correlations in the smallest scale bins due to the masking correction.
}\label{fig:correlation}
\end{figure*}

In addition to measurements of power spectrum, our statistical analysis requires a covariance matrix.
The full covariance matrix consists of \(n_\textup{bins}\cdot(n_\textup{bins}+1)/2\) independent entries where \(n_\textup{bins}\simeq 15\) is the number of band powers
of the wavenumber \(k\).  For a dataset like ours that consists of only \(n_\textup{QSO}\sim 10\) quasars, this estimate will be extremely noisy.
\citet{mcdonald2000ObservedProbabilityDistribution} did tests on bootstrapped covariance matrices based on spectra that were subdivided into 5 spectral chunks (which then were treated as independent data in the same redshift bin) and concluded that there was still significant statistical uncertainties on the estimates of individual covariance matrix entries.
Also \citet{irsic2016Lymanalphaforestpower} tested the bootstrap covariance estimation technique on models with different amounts of skewers. When using only 100 model skewers, which is comparable to the size of their dataset,
noise in their correlation matrix is still clearly visible.
They also provide a correlation matrix of their measurement that looks similar to this model estimate.
As our typical redshift bin has less data available than theirs and as our data is additionally masked for metal absorption, we do not believe that we can
estimate reliable covariance matrices from our data (see \autoref{fig:correlation} and discussion below).
To circumvent this problem we use a hybrid approach \citep[in a similar way as][]{Lidz2006TighteningConstraints}
and only measure the diagonal values \(\sigma_\mathrm{data}^2\) of the covariance from the data itself while computing the off-diagonal correlations \(R_\mathrm{ij}\) from simulations (see \autoref{sec:dm-skewers}) for which we can obtain sufficient statistics:
\begin{align}
R_\mathrm{ij}=\frac{\langle(P_\mathrm{m,single}(k_\mathrm{i}) - P_\mathrm{m}(k_\mathrm{i}))
(P_\mathrm{m,single}(k_\mathrm{j})-P_\mathrm{m}(k_\mathrm{j}))\rangle}{\sigma_\mathrm{m}(k_\mathrm{i}) \sigma_\mathrm{m}(k_\mathrm{j})}
\label{eq:corrmat}
\end{align}
where \(P_\mathrm{m,single}\) is the power estimated from a single model skewer, \(P_m\) is its mean over all model skewers, \(\sigma_\mathrm{m}\) its standard deviation and the average \(\langle\dots\rangle\) is performed over all model skewers.
The covariance matrix is then computed as:
\begin{equation}
C_\mathrm{ij}=\sigma_\mathrm{data}(k_\mathrm{i}) \sigma_\mathrm{data}(k_\mathrm{j}) R_\mathrm{ij}.
\end{equation}

To estimate the data uncertainties \(\sigma_\mathrm{data}\) we use a bootstrapping technique resampling the data.
For this we draw 1000 random sets of quasars with replacement from those contributing to any given redshift bin, and for each compute the power spectrum
using \autoref{eq:power_definition}.
For the correlation matrix \(R\), we use the same mock spectra generated with our forward modeling procedure (the `forward models' in \autoref{sec:fwmodel}) to determine the correlation matrix according to
\autoref{eq:corrmat}. No bootstrapping was performed, since for the mocks we have access to different forward modeling realizations of the same dataset. Note that we have a correlation matrix for
each model, because the shape of the power spectrum impacts the correlations leading to stronger correlations for modes on scales smaller than the power spectrum cutoff.

In \autoref{fig:correlation} we show the correlation matrix for one of our redshift bins determined
from the model with parameters closest to our best-fit at this redshift.  The typical correlation between neighboring bins is \(\sim15\%\) and decreases strongly for bins further apart.
For comparison,  we also compute the correlation matrix from our dataset at this redshift as well. This correlation matrix is noisy, and is not used anywhere else in our analysis but is shown here for the sake of illustration. However,
one sees that it qualitatively agrees well with our simulated correlation matrix, validating our approach.
We also show the correlations of a perfect model and comparing to the forward model we can see that the additional masking added significant correlations on the \(20\%\) level between non-neighboring \(k\)-bins.

We also tested our bootstrap estimation of the diagonal elements of the covariance using random simulated datasets with a size comparable to our measurement dataset that were drawn from our full set of models without replacement.
We found that the variance of the power spectrum determined from the ensemble of simulated datasets was in good agreement with the bootstrap estimate obtained from a single mock dataset.
Therefore we are confident that our bootstrap estimates of the diagonal elements of the covariance are converged and reflect the actual uncertainties of the power spectrum.

\subsection{Fast Emulation of Model Power Spectra} \label{sec:emulator}

To be able to fit the power spectrum measurement we need to create a model for the power in the full region of interest for thermal parameters (\(T_0, \gamma,\lambda_P\)) and
the mean transmission of the forest \(\bar{F}\).
As simulations are relatively expensive there is no way to run a full
simulation for every combination of thermal parameters at which the
power spectrum needs to be evaluated during the fitting process.
Therefore we generate a grid of simulations with different thermal
parameters (using our semi-numerical \ac{DM} only simulations)
and adopt a fast emulation procedure to compute the power spectrum for
parameters between the grid points \citep[in a similar way to what is
  used for the cosmic calibration framework
  by][]{heitmann2006Cosmiccalibration, heitmann2009CoyoteUniverseII, Heitmann2013COYOTEUNIVERSE, habib2007CosmicCalibrationConstraints}.  Following the procedure
described in \autoref{sec:dm-skewers}, we generated Ly$\alpha$ forest skewers
for a grid of thermal parameters (\(T_0, \gamma, \lambda_P\)) and mean
flux values \(\bar{F}\). We computed the power spectrum from perfect
skewers generated from these models, as well as for mock data run
through the forward modeling pipeline described in \autoref{sec:fwmodel}.

For both sets of power spectra (perfect and forward-modeled) we perform a \ac{PCA} of the full set of model power spectra. As result the power
at any point \(\theta=\{T_0,\gamma,\lambda_P,\bar{F}\}\) in our model grid can be written:
\begin{equation}
\label{eq:PCA}
 P(k, \theta)=\sum_{i} \omega_{i}(\theta) \Phi_{i}(k),
\end{equation}
where both the eigenvectors \(\Phi_{i}(k)\) and the \ac{PCA} weights \(\omega_{i}(\theta)\) are a result of the \ac{PCA} decomposition.
Note that the \ac{PCA} weights depend on the gridpoints in our model parameter space, and given a suitable interpolation scheme, they can be
used to generate model predictions at any point in this space. Along these lines, and following \citet{habib2007CosmicCalibrationConstraints}, the \ac{PCA} weights \(\omega_{i}\) are then used as an input to train a
Gaussian process interpolation scheme.  From this Gaussian process we can later evaluate new weights for parameter combinations that lie between or original gridpoints,
allowing us to evaluate the power at any location. For more details on this approach see a companion paper to this analysis \citep{Walther2017EmulatingLyman-alpha-forest} or e.g. \citet{rorai2013NewMethodDirectly} and \citet{habib2007CosmicCalibrationConstraints}.
The Gaussian process interpolation uses a squared exponential kernel with smoothing lengths chosen to be larger than the separation between model grid points of the parameter space.
To speed up computations only the first 9 principal components are actually used for the analysis.
We found that the additional errors due to discarding the higher order components is less than 1\%.

In general the same approach can be used to emulate models from hydrodynamical simulations, but this means that a separate hydrodynamical simulation must be run for each parameter combination in the grid which is far more costly than the approach we chose for this work.
Nevertheless, we will use an emulator based on full hydrodynamic simulations to infer thermal parameters from our power spectrum measurements in a companion paper
\citep{waltherinprepConstraintsthermalevolution}.

\subsection{Parameter Exploration}\label{sec:param_estim}

To explore the parameter space and fit the measured data power spectra from both BOSS and high-resolution datasets we use a Bayesian \ac{MCMC} approach with a Gaussian multivariate likelihood:
\begin{align}
\mathcal{L}\equiv&P(\mathrm{data}|\mathrm{model})\\
\propto&\prod_\mathrm{datasets}\frac{1}{\sqrt{\det(C)}}\exp\left(-\frac{\mathbf{\Delta}^\mathrm{T} C^{-1} \mathbf{\Delta}}{2}\right) \nonumber\\
\mathbf{\Delta}=&\mathbf{P}_\mathrm{data}-\mathbf{P}_\mathrm{model} \nonumber,
\end{align}
where \(\mathbf{P}_\mathrm{data}\) and \(C\) are the power spectrum and covariance matrix obtained for any  dataset, and the product is over all datasets considered (i.e. BOSS and our
high-resolution measurement, but including more than two datasets would just add factors to this product).

For the high-resolution dataset \(C^{-1}\) is estimated using our
hybrid approach (see \autoref{sec:covmat}) for each model on the parameter grid and therefore is
a model dependent quantity.  Inside the likelihood computation we use
nearest neighbor interpolation in model parameter space to obtain an inverse covariance
estimate for each combination of model parameters (which is then just
the matrix at the closest model grid point).  For the covariance matrix of the
BOSS dataset we use the tabulated values from
\citetalias{palanque-delabrouille2013onedimensionalLy} which are directly measured from the data, and are therefore model independent.
The model power \(\mathbf{P}_\mathrm{model}\) is determined for any parameter combination using our emulation procedure. Note that we actually have two distinct emulators, one for
the high-resolution data (using the full forward-modeling procedure), and one for the BOSS data (using our perfect simulation skewers).

It is well known that correlated absorption of hydrogen \ac{Lya} and \SiIII{} at \(\SI{1206.5}{\angstrom}\) leads to a bump in the \ac{Lya} forest flux correlation function \(\xi (\Delta v)\) at \(\Delta v=\SI{2271}{\kms}\) \citepalias[\citealt{mcdonald2006LyForestPower};][]{palanque-delabrouille2013onedimensionalLy}\footnote{Note that this correlation should generally be weaker in our high-resolution dataset as we masked some of the \SiIII{} absorption.}. This imprints wiggles on the power spectrum with separation \(\Delta k=2\pi/\Delta v=\SI{0.0028}{\skm}\). Following \citet{mcdonald2006LyForestPower}, we model this contamination with a multiplicative correction to the power
\begin{equation}
P (k) = (1+a_\mathSiIII^2+2a\cos(\Delta v \, k))P_\mathHI{}(k),
\label{eq:siIII}
\end{equation}
with \(a_\mathSiIII\) being a free nuisance parameter for the strength of the correlation.
In previous works this was typically expressed as \(a_\mathSiIII=f_\mathSiIII/(1-\bar{F})\) with \(f_\mathSiIII\) being a redshift independent quantity that was fit using the entire
dataset. We adopt this same parameterization but opt to fit for a unique value of \(f_\mathSiIII\) at each redshift.  Therefore we have five
free parameters \(T_0,\gamma,\lambda_J,\bar{F},f_{\mathSiIII}\) to fit to our power spectrum measurements in each redshift bin.

We used the publicly available \ac{MCMC} package \texttt{emcee} \citep{foreman-mackey2013emceeMCMChammer} based on an affine invariant ensemble sampler \citep{goodman2010Ensemblesamplersaffine} to sample the posterior distribution:
\begin{equation}
P(\mathrm{model}|\mathrm{data})=\frac{P(\mathrm{data}|\mathrm{model})P(\mathrm{model})}{P(\mathrm{data})}
\end{equation}
with \(P(\mathrm{data}|\mathrm{model})\) being the combined likelihood of both datasets for a given set of model parameters, and \(P(\mathrm{model})\) being the prior on that combination of parameters.
We assume that the priors on parameters are independent, therefore \(P(\mathrm{model})\) is just the product of the individual priors for each parameter.
For the thermal parameters we adopt flat priors on \(\gamma\) in the range \(0.5<\gamma<2.1\), \(\log(\lambda_J)\) in the range \(\SI{22}{\ckpc}<\lambda_J<\SI{150}{\ckpc}\) and \(\log(T_0)\) in the range \(\SI{3000}{K}<T_0<\SI{20000}{K}\).
For the mean flux we adopt Gaussian priors based on the most recent measurements of
\citet{becker2013refinedmeasurementmean}, \citet{faucher-giguere2008DirectPrecisionMeasurement} and \citet{kirkman2005HIopacityintergalactic} (depending on redshift) for \(\bar{F}\). For the \SiIII{} correlations  we used a Gaussian prior defined by the best-fit value in \(f_\mathSiIII\) and \(1\sigma\) region of \citet{palanque-delabrouille2013onedimensionalLy}.
All priors for all redshifts are listed in \autoref{tab:priors}.
Note that as long as a good fit to the data can be obtained (i.e. the model is sufficiently flexible) the exact values of thermal parameters (and therefore also the exact priors chosen) don't matter in the context of this work as the fits are only used to correct out the window function as we discuss in the next subsection.
In a companion paper doing thermal parameter analysis based on hydrodynamical simulations we will loosen the priors on both \(\bar{F}\) and \(f_\mathSiIII\) to not bias our results.
Fits were always performed for individual redshifts and no correlation between thermal properties at different redshifts was assumed.

\subsection{The Raw Power Spectrum and Window Function Correction}
\label{sec:raw-measurement}

\begin{figure*}
\centering
 \plottwo{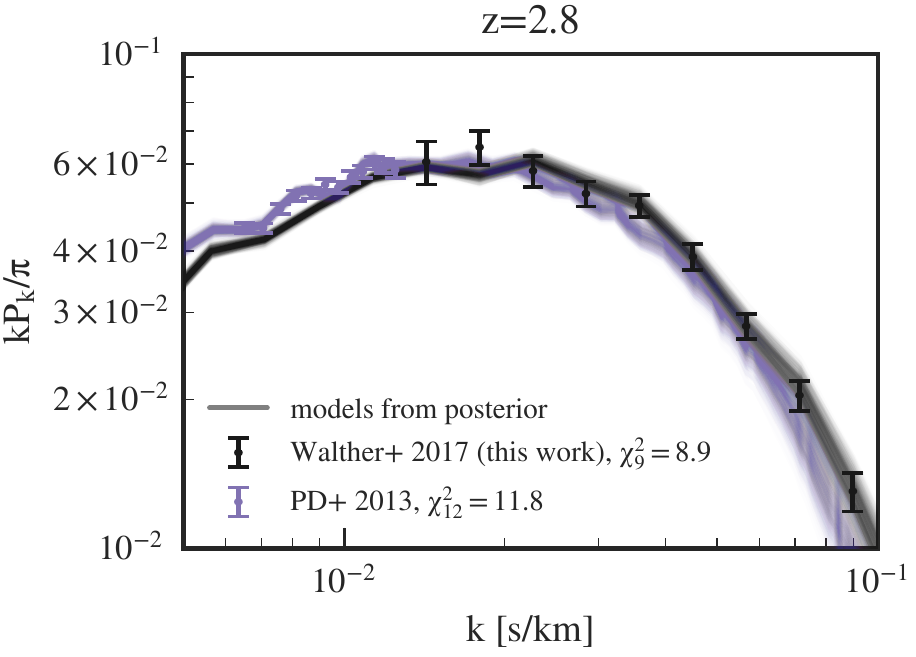}{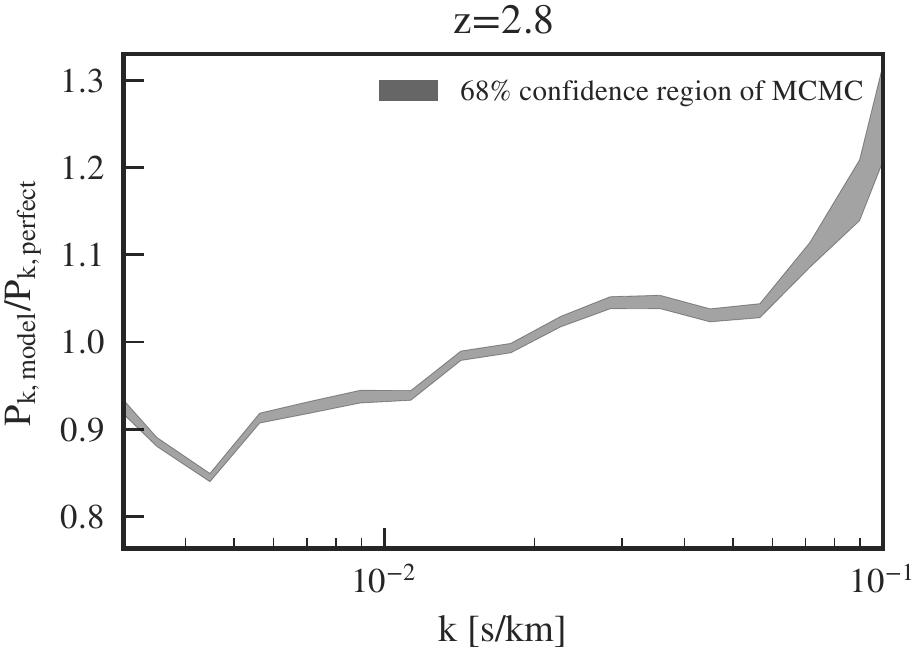}
\caption{Left:
The measured power spectrum at redshift \(z=2.8\) compared to the best-fit.
The black points show the raw power of the high-resolution data and are fitted with mock observations including the same noise and masking as in the data.
The purple points are showing the \citetalias{palanque-delabrouille2013onedimensionalLy} measurement including their metal, noise and resolution corrections.
These are then fitted with models generated from noiseless, non-masked skewers.
Lines show models randomly drawn from the posterior MCMC-chain.\\
Right:
The window function estimate from a comparison between the power from noiseless, mask-free models and the full forward-modeled mock observations.
The band shows the 68 \% contour for models drawn from the posterior MCMC-chain.
The median is applied as a correction factor to the raw data.
The width of the band is propagated into the data errorbars.}\label{fig:dm_data_compare}
\end{figure*}

The impact of masking on the power depends on the underlying shape of the power spectrum.  We use the fits to the underlying power based on our DM models to determine the impact
of the window function on the shape of the power spectrum
We emphasize here that we use this window-function corrected power mainly for visualization purposes
only as the covariance in the measurement is subtly changed by the correction.
While we do provide an approximate covariance matrix in \autoref{sec:dataproducts}, ideally a full forward model of the power spectrum should be used when doing parameter estimations.

Our measurement, emulation, and window function procedure are illustrated in \autoref{fig:dm_data_compare}.
On the left side we show a comparison between our raw data power spectrum (i.e. before window function correction) and the BOSS measurement as points.
After applying our fitting procedure we obtain an MCMC chain from which we can draw parameter combinations \(\Theta=\{T_0,\lambda_P,\gamma,\bar{F},f_\mathSiIII\}\) that are compatible with the data.
Feeding random draws from this chain into our emulator routines for both the forward model and the perfect model produces the black and purple bands.
We can see that these provide good fits to our dataset and the BOSS dataset, respectively.
For a single draw \(\Theta\) from the posterior, we can then measure a window function correction \(f_\mathrm{window}\) using both emulators as:
\begin{equation}
f_\mathrm{window}(k)=\frac{P_{\mathrm{forward\,model}}(k,\Theta)}{P_{\mathrm{perfect\,model}}(k,\Theta)}.
\end{equation}
The gray bands on the right side of \autoref{fig:dm_data_compare} show the resulting \(16\%\) and \(84\%\) quantiles of this quantity.
The dominant effect of windowing our data is a strong increase in power on the small scale (\(k\gtrsim\SI{0.07}{\skm}\)) end of the measurement.

To extract the underlying (i.e. window function corrected) power spectrum from our raw measurement we proceed as follows.
We generate 1000 random draws from a multivariate normal distribution with a mean given by the raw \(P(k)\)
measurement (which includes the effect of the window function) and a covariance matrix defined in \autoref{eq:corrmat} (based on both the raw measurement and the best-fit model).
We then obtain the window function correction \(f_\mathrm{window}(k)\) for 1000 draws from the posterior.
By multiplying each \(P(k)\) draw with each \(f_\mathrm{window}(k)\) draw we obtain samples of the distribution describing the window corrected measurement.
We take the mean of these samples to represent our window corrected power,
and the covariance of these samples gives the respective window corrected
covariance.
We show the resulting correlation matrix applying this procedure at \(z=2.8\) in \autoref{fig:correlation} (lower right panel).
Compared to the uncorrected forward model we can see additional correlation at the smallest scales (\(k\gtrsim\SI{0.07}{\skm}\)) where the influence of the window correction is strongest.
{The window function corrected measurements and correlation matrices are tabulated in \autoref{sec:dataproducts} and can be used for comparison with other datasets as well as model fitting. We also provide MCMC chains of \(f_\mathrm{window}\) based on our forward models to facilitate reproduction of our measurements.}

\section{A New Power Spectrum Measurement} \label{sec:results}
In this section we present our final window-function corrected power spectrum measurements over the redshift range \(1.8 \le z \le 3.4\). First, we discuss the impact of metal line contamination on the power spectrum measurement.
Then we compare  our power spectrum measurement to the lower-\(k\) measurements from BOSS as well as the XQ-100 dataset,
which represent the state-of-the-art from low- and medium-resolution data. Finally, we compare our results to previous
high-resolution measurements.

\subsection{The Final Power Spectrum Measurement and the Effect of Metals on the Data}

\label{sec:metalcomp}
\begin{figure*}
\centering
 \plotone{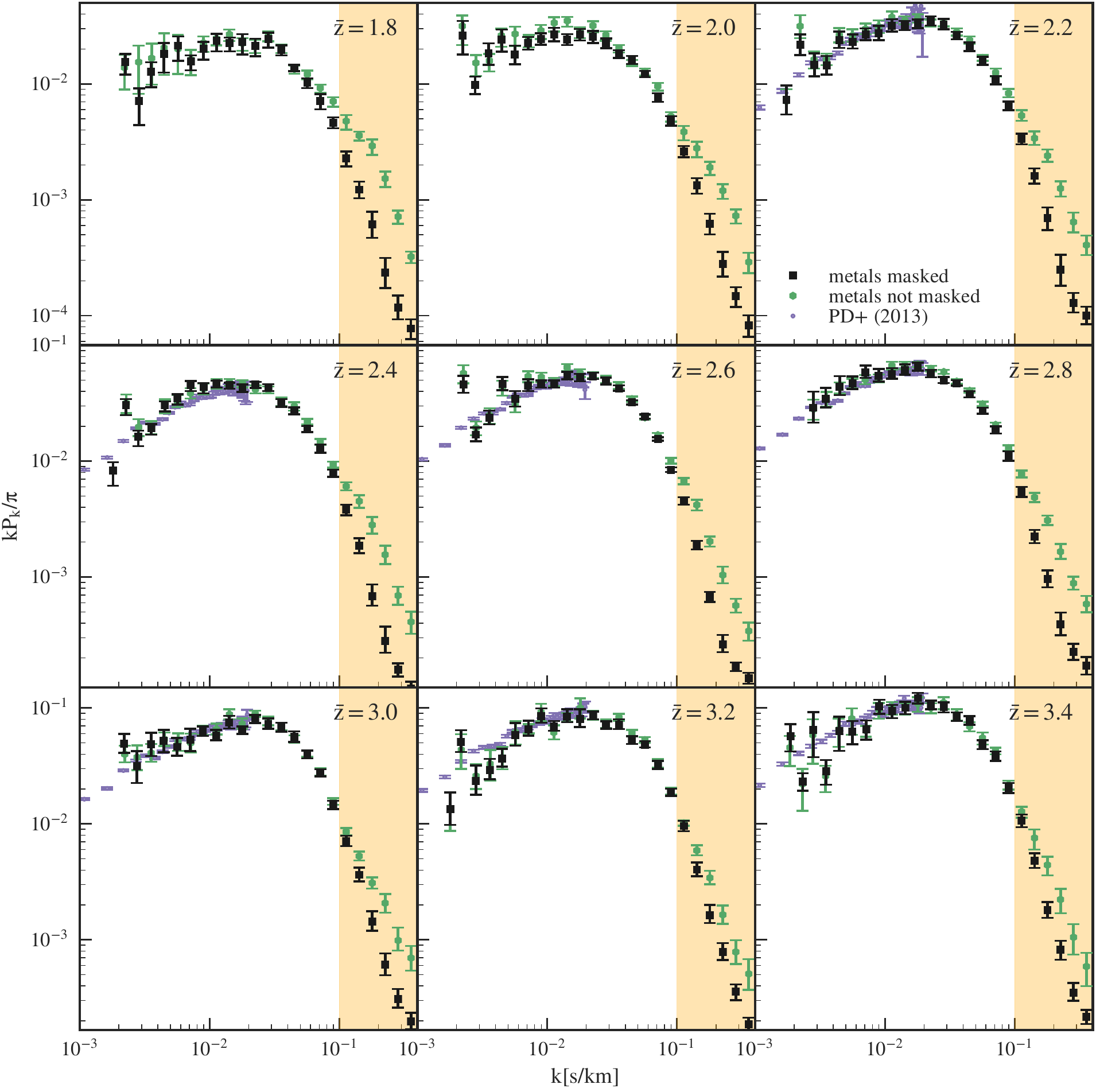}
\caption{Our power spectrum measurement of spectra with masked metals (black squares) and with metals left inside the forest (green hexagons) as well as the BOSS measurement (purple errorbars).
Both high-resolution measurements are corrected for their respective window function.
In most redshift bins contamination due to metal absorption clearly leads to an increased small scale (large \(k\)) power which would bias a potential temperature measurement towards lower temperatures.
The orange region shows the region excluded in our further analysis as this effect gets far larger than our statistical uncertainties.
On larger scales power is removed as well when masking the metal lines leading to an overall better agreement with the BOSS measurement.}\label{fig:powerspec_metal}
\end{figure*}

In previous work the effect of metals on the small scale power spectrum has been largely ignored.
While \citet{mcdonald2000ObservedProbabilityDistribution} computed the power spectrum on a dataset with and without metals masked  and also considered the effect of masking on the power spectrum, they did not combine both results to see the net-effect of metal contamination.
However, they did note that especially for \(k>\SI{0.1}{\skm}\) the effect of metals as well as noise becomes too strong for their measurement to be usable.
Motivated by this conclusion, later measurements by \citet{croft2002PreciseMeasurementMatter}, \citet{kim2004powerspectrumflux} and \citet{viel2008HowColdCold,viel2013Warmdarkmatter} also ignored these small scale (larger \(k\)) modes.
For lower resolution measurements using SDSS \citep{mcdonald2006LyForestPower} or BOSS \citepalias{palanque-delabrouille2013onedimensionalLy} spectra, these
modes are well above the resolution limit of the data,
and this has not been an issue.
Instead, for those measurements the power in metals is estimated from lower redshift data (using a spectral range redwards of the \ac{Lya} forest).
As this procedure cannot treat metal lines that are always inside the forest large scale correlations between \SiIII{} and \ac{Lya} are the dominant effect of metal contamination in this case.

In \autoref{fig:powerspec_metal} we show a comparison between our
power spectrum measurement applying the metal masking procedure
described in \autoref{sec:mask} and performing the analysis without
masking metals.  Both measurements have been noise subtracted
following the discussion in \autoref{sec:ps-measurement-method} and
corrected for their respective window functions according to \autoref{sec:raw-measurement}. We also show the BOSS measurement as a comparison, for \(z > 2.2\) where those measurements exist.

It is clear that particularly at small scales (\(k\gtrsim \SI{0.1}{\skm}\)) metal lines
significantly contribute to the measured power spectrum.
This increase in small scale power in general leads
to an underestimation of the small-scale thermal cutoff, and naively
fitting this metal-contaminated power would result in lower overall
IGM temperatures (i.e. more small-scale structure and hence less thermal broadening and/or pressure
smoothing).  For all further analysis we therefore
use the metal-masked power (black dots in the figure), and our model fitting
is conservatively restricted to modes with \(k<\SI{0.1}{\skm}\), where the impact of metal line contamination is
relatively weak, and hence our metal-masked power is relatively insensitive to the fact that we may not have masked
all of the metals (see discussion in \autoref{sec:dataprep}).

\autoref{fig:powerspec_metal} also indicates that the impact of metal-line contamination is not restricted to small-scales. Multiple ionic metal-line transitions are typically associated with a given absorber redshift, and because the velocity separations are thousands of \(\si{\kms}\), these large-scale velocity correlations can impact the power spectrum on large-scales (low-\(k\)) as well.
Masking metals in the data at e.g. z=2.6 decreases this effect lowering the power spectrum and therefore increases the agreement with the \citetalias{palanque-delabrouille2013onedimensionalLy} measurement.

To summarize, metal-line contamination of the power spectrum does not
significantly impact our results, given that we mask the metals and
conservatively restrict our power spectrum fits to low
\(k<\SI{0.1}{\skm}\) where the impact of residual (i.e. metals we
missed in our masking) metal-line contamination should be negligible.
In principle, a more careful treatment of metal-lines (i.e. via improved masking, or
forward modeling the metals, or subtracting the red-side metal-line power) could allow
one to access even smaller scales (higher \(k\)), although this would also require a very
careful treatment of the noise and instrumental resolution whose effect also increases for smaller scales (larger \(k\)).

\subsection{Comparison to Previous Low and Medium Resolution Measurements}
\label{sec:final-measurement}

\begin{figure*}
\centering
\plotone{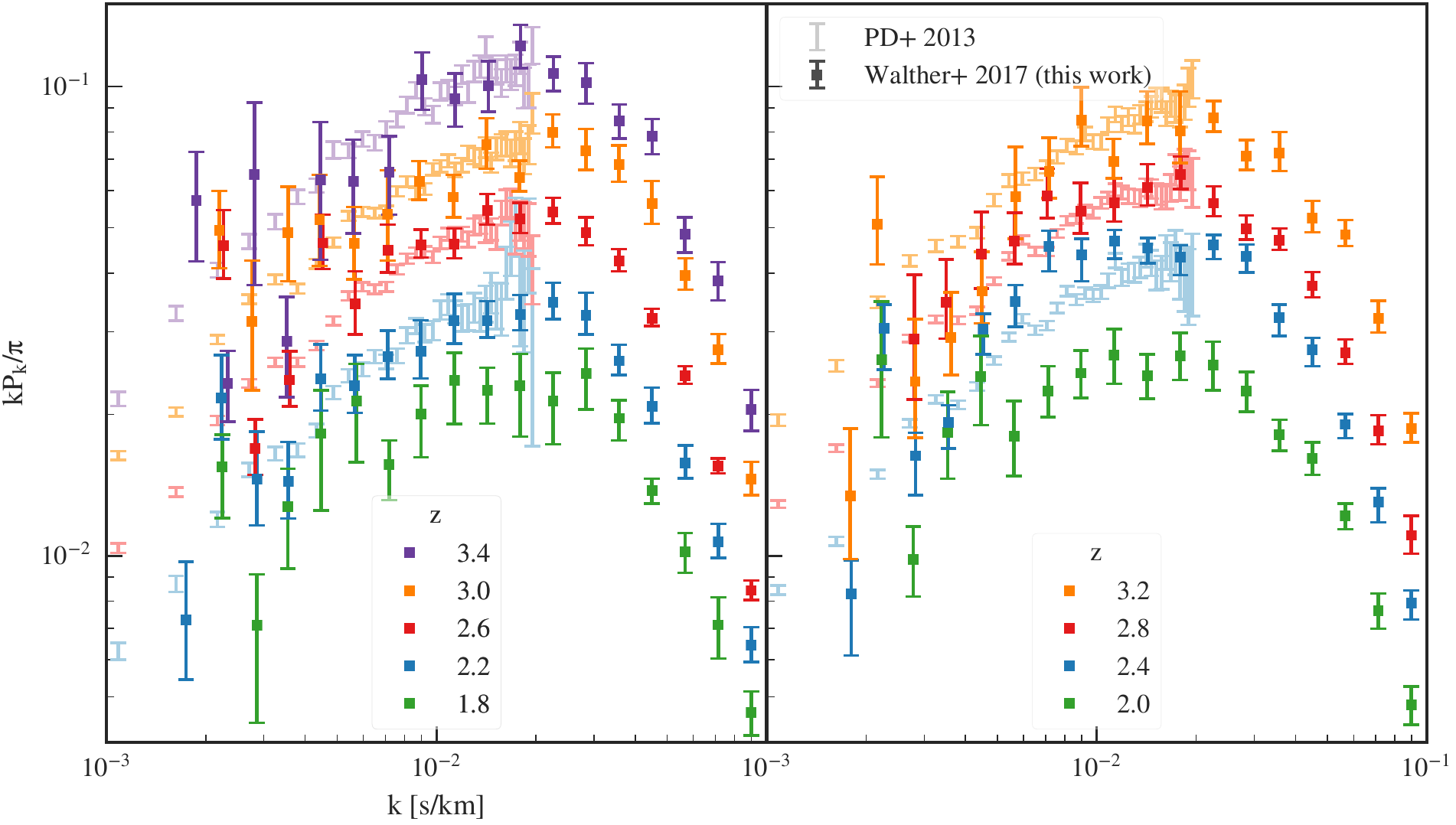}
\caption{Our new measurement of the power spectrum (dark squares) for \(1.8\leq\bar{z}\leq 3.4\) (different colors), metal lines were masked our analysis and the power introduced by masking was removed using forward modeling of our measurement (see \autoref{sec:fwmodel}). Also shown are the measurements of \citetalias{palanque-delabrouille2013onedimensionalLy} (BOSS, bright errorbars on the left). We can see that at most overlapping redshifts there is good agreement except for a mild disagreement with BOSS on large scales (small \(k\)) for \(z\sim2.4\).}\label{fig:powerspec_allz}
\end{figure*}

In \autoref{fig:powerspec_allz} we show our new metal and window
corrected measurement of the high-resolution power spectrum compared
to the \ac{BOSS} measurement from
\citetalias{palanque-delabrouille2013onedimensionalLy}. Note that different
power spectrum bins are correlated and the errorbars only reflect the
diagonal elements of the covariance matrix.  Where both measurements
overlap (\(k\lesssim \SI{2e-2}{\skm}\), \(2.2<z<3.4\))  we find good agreement with the BOSS power
spectrum (which has a typical accuracy of \(\sim 2\%\)) within our
errors at modes \(k>\SI{0.01}{\skm}\) and the agreement is generally
good for larger scale (lower-\(k\)) modes as well. The discrepancies
at low-$k$ are most likely due to the fact that our measurements are
rather noisy for small wavenumbers\footnote{This is a result of two
factors: 1) the density of modes for a one-dimensional power spectrum
measurement is uniform, and so for linearly spaced bins there are
equal numbers of modes at each $k$. However, our $k$-bins are
logarithmically spaced, so in general our error bars are larger in a
relative sense at lower-$k$. 2) For some of our spectra either
limited spectral coverage, or our masking procedure tend to reduce the
amount of path length available for measuring the largest scale modes,
making them more noisy.}.

\begin{figure*}
\centering
\plotone{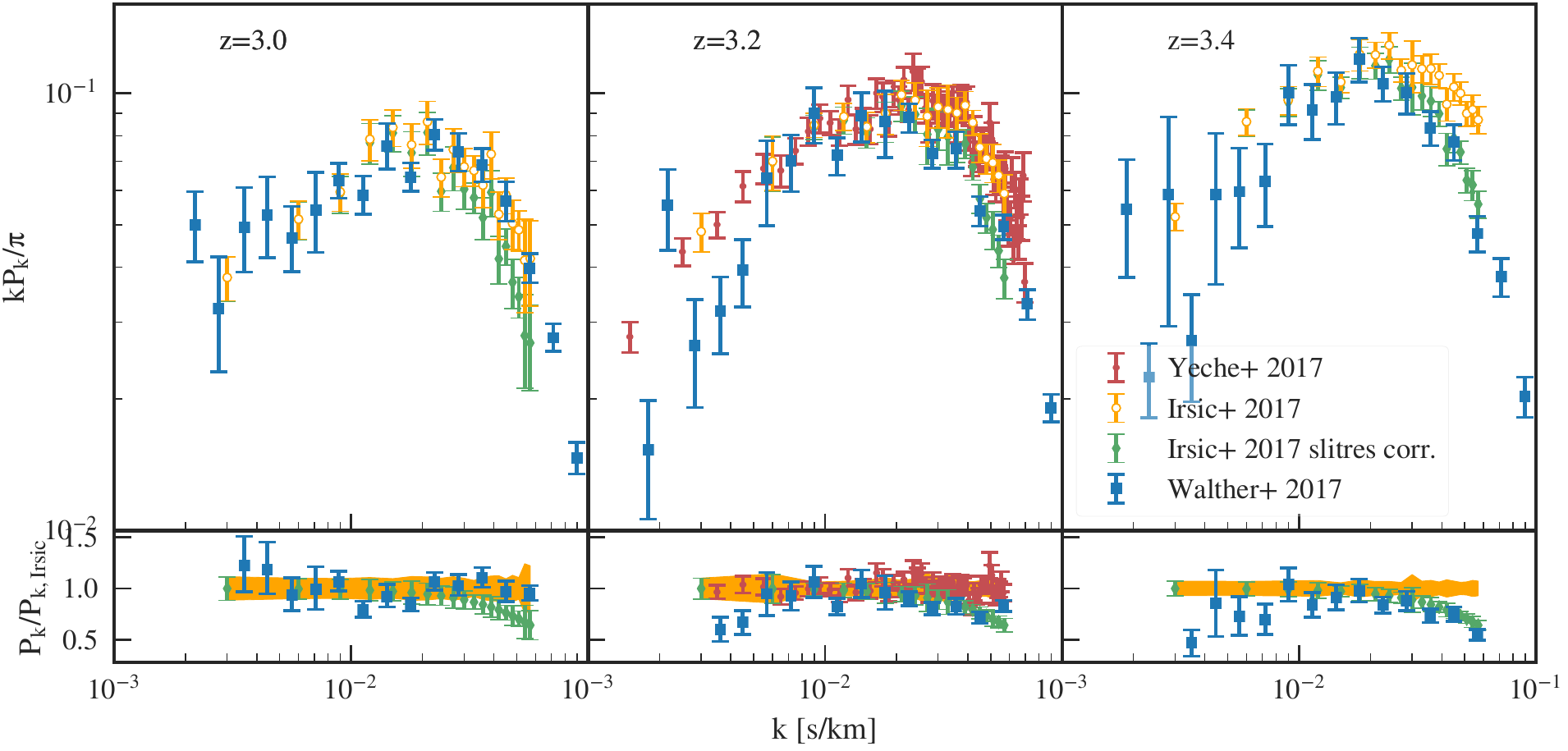}
\caption{Our new measurement of the power spectrum (blue squares) for \(3.0\leq\bar{z}\leq3.4\) as in \autoref{fig:powerspec_allz} compared to the {measurements from XQ-100 by \citet{irsic2016Lymanalphaforestpower} (orange open circles) and \citet{yeche2017Constraintsneutrinomasses} (dark red dots)}. To address the disagreement at small scales (high \(k\)) for \(z\geq3.2\) between our high-resolution and the XQ-100 data, we also show the XQ-100 data {of \citet{irsic2016Lymanalphaforestpower}} assuming a different resolution correction (see main text for details) by using a distribution of seeings and assuming an underestimation of the X-SHOOTER slit-resolution (green diamonds).
The bottom panel shows the same comparison, but normalized by the (untreated) Irsic+2017 measurement. The quoted \citet{irsic2016Lymanalphaforestpower} measurement errors are shown as an orange band.}\label{fig:powerspec_Irsic}
\end{figure*}

We also compare to the recent results of \citet{irsic2016Lymanalphaforestpower} {and \citet{yeche2017Constraintsneutrinomasses}} based on the XQ-100 dataset \citep{lopez2016XQ100legacysurvey}. Specifically, \citet{irsic2016Lymanalphaforestpower} measured the power at \(3.0 \leq z \leq 4.2\), and their redshift bins at $z=3.0$,$3.2$, and $3.4$ match those used in our analysis {(\citet{yeche2017Constraintsneutrinomasses} chose to use broader bins and their \(z=3.2\) bin can be compared to our results)}. Those are compared in \autoref{fig:powerspec_Irsic}.

While the agreement between our measurements at \(z=3\) is good , at $z=3.2$ and $z=3.4$ \citet{irsic2016Lymanalphaforestpower}
measures \(\sim 10\%\) to \(50\%\) higher power than we do at small scales \(k\gtrsim \SI{0.02}{\skm}\), which is clearly statistically significant given the error bars for the respective datasets.
Note however that this disagreement is restricted to high-$k$, and we agree well with \citet{irsic2016Lymanalphaforestpower} at intermediate $k$ in all redshift bins.

While it is difficult to be certain about the source of this discrepancy, given the different methodology used for measuring the power, the \(k\) dependence of this disagreement provides a very important clue.
Note that X-SHOOTER data is significantly lower resolution than the dataset presented here and the resolution corrections become significant at higher-\(k\).

Possible uncertainties of spectroscopic resolution can come from several sources: 1) as X-SHOOTER is a slit-spectrograph its resolution is seeing dependent, so the seeing itself needs to be accurately determined to get an accurate resolution, 2) seeing changes between different observations and correcting assuming a constant resolution across the dataset might bias the measurement (see \autoref{sec:reseffects} for more details on those points), 3) the UVB slit resolution quoted for X-SHOOTER might be significantly underestimated (see \autoref{sec:slitres}).
Because of those problems, a correction based on the nominal slit resolution can overproduce small scale power in the data analysis.
However, \citet{irsic2016Lymanalphaforestpower} already used a higher resolution than provided in the XQ-100 data release paper when performing his corrections\footnote{They used a FWHM resolution of \(\SI{50}{\kms}\) instead of the nominal \(\mathrm{c}/R=\SI{69}{\kms}\) where \(R\) is the quoted resolving power of the X-SHOOTER spectrograph. Communication with the main author revealed that this value was obtained by visual inspection of the raw science data before the official values were available in {what he claims to be} a similar procedure as in the \citet{yeche2017Constraintsneutrinomasses} analysis.}.
Using the \citet{irsic2016Lymanalphaforestpower} value for the resolution leads to agreement between both XQ-100 power spectrum analyses by \citet{irsic2016Lymanalphaforestpower}
and the independent determination of the power spectrum from the
same dataset by
\citet{yeche2017Constraintsneutrinomasses}. \citet{yeche2017Constraintsneutrinomasses} determined
the spectral resolution by assuming the XQ-100 resolution, and estimating the
seeing on object by object basis analyzing the width of the \ac{Lya} forest in the spacial direction of the 2d-spectra.
{While this approach should remove the sensitivity towards seeing, it cannot tackle a potentially misestimated slit resolution or changes of this quantity along the spectral arm. \citet{yeche2017Constraintsneutrinomasses} also chose to not combine the different spectral arms. Thereby, there shouldn't be a strong change of resolution when reaching the overlap region between both spectral arms. Thus, inside each redshift bin their data should be more homogeneous regarding resolution. Nevertheless, in the end both measurements seem to give essentially the same result.}

The influence of resolution errors on the resolution correction factor \(W_R^2\) (see \autoref{eq:res_correction}) can be found by simple error propagation:
\begin{equation}
\begin{aligned}
\Delta \ln W_R^2 &= -k^2 \Delta (R^2)\\
&\simeq -2k^2 R^2 \Delta \ln (R)
\end{aligned}
\end{equation}
and propagates to an error on the estimated power spectrum \(P\) as:
\begin{equation}
\Delta \ln P = -\Delta \ln W_R^2.
\end{equation}
Assuming the nominal resolving power of 4350  \citep[according to][listing a slightly higher value than the ESO Instrument description]{lopez2016XQ100legacysurvey} using the X-SHOOTER \(\SI{1}{\arcsecond}\) slit on the UVB arm as a worst case scenario a spectrum with \(10\%\) higher resolution than assumed when performing the correction will lead to a \(\sim 45\%\) (\(28\%\) assuming a resolving power of 5350) overestimate of the power in the resolution corrected measurement.
As our own power spectrum measurement is based on \(\sim10\) times higher resolution data a comparable error in the knowledge of the resolution
will have a \(\sim100\) times smaller effect (so \(\sim 0.4\%\) at \(k=\SI{0.05}{\skm}\) and \(\sim 1.5\%\) at \(k=\SI{0.1}{\skm}\)). Therefore lack of
knowledge of the precise resolution of the spectrograph a significant concern
for the X-SHOOTER measurement, but can be safely ignored for our study.

To determine the influence of possible errors in the resolution estimates due to points 1 and 2 we divided out the resolution correction
from the \citet{irsic2016Lymanalphaforestpower} power spectrum measurement and corrected using different assumptions. First, we used the nominal slit resolution \(R=4350\) from the X-SHOOTER spectrograph, generated Gaussian distributed seeing values (with a mean of \(\SI{0.75}{\arcsec}\) and FWHM of \(\SI{0.2}{\arcsec}\) similar to the distribution shown in \citealt{yeche2017Constraintsneutrinomasses}) for each we computed the resolution correction \(W_R^2\) according to \autoref{eq:res_correction} and obtained the mean correction which we then used as the new resolution correction.
This gives basically identical results (that we don't show in our comparison figure for clarity) to the original \citet{irsic2016Lymanalphaforestpower} measurement showing that the seeing estimate used for their measurement is in agreement with the distribution determined by \citet{yeche2017Constraintsneutrinomasses} and cannot be the reason for the disagreement with our measurement.
In addition we also generate a measurement assuming a higher slit resolution of \(R=5350\) (due to point 3 and in agreement with measurements based on calibration spectra, see \autoref{sec:slitres})
and otherwise performing the same analysis.
This comparison is shown as red diamonds in \autoref{fig:powerspec_Irsic} and we can see that the agreement between our high-resolution measurement and XQ-100 at \(z=3.2\) and \(z=3.4\) is good in this case.
However, the agreement in the \(z=3.0\) bin without assuming a different X-SHOOTER resolution correction is unclear, but might be due to possible variations in the resolution between Echelle orders.
Note that for the other spectral arms (that cover the \ac{Lya} forest for \(z>3.6\)) this is a less severe problem as data from those is intrinsically higher resolution.
Also note that for those arms one can in principle obtain the resolution of the science observation from the width of telluric absorption lines \citep[see][]{Bosman2017deepsearch}, but those are rare in the UVB arm.
This can also explain the agreement between XQ-100 measurements and older high-resolution data by \citet{mcdonald2000ObservedProbabilityDistribution} at \(z\sim 3.8\) (but as the redshift bins of both measurements are significantly different, this comparison is tricky) and \citet{viel2013Warmdarkmatter} at \(z=4.2\).

Because of the severe impact the resolution correction can have on the XQ-100 power spectrum measurement we caution against using the smallest scales (\(k>\SI{2e-2}{\skm}\)) of this measurement (at least for the lowest redshift bins) for parameter studies. This is especially true for measurements that rely on the accurate determination of the power spectrum cutoff, like e.g. determining the thermal state of the IGM, or the nature of dark matter \citep[e.g.][]{Irsic2017Newconstraints,irsic2017Firstconstraintsfuzzy,baur2017ConstraintsLybackslash}. Although to be fair, for the latter most of the sensitivity comes
from the higher redshift (\(z \gtrsim 4\)) bins where the resolution of X-SHOOTER is higher and additionally high-resolution ($R \simeq 50,000$) data from \citet{viel2013Warmdarkmatter} are used.

In \autoref{fig:metal-mask-lines} we can see, that our metal removal and masking correction techniques do not change the data hugely (the difference is covered by our error bars) at the redshifts and scales where we disagree with \citet{irsic2016Lymanalphaforestpower}. Additionally, the changes due to metal masking do not exhibit the same shape as the discrepancy. Finally, we also checked the raw power spectra not corrected for any masking and could not get a small-scale power as high as in the \citet{irsic2016Lymanalphaforestpower} result. We are therefore confident of metal masking and window correction not being the reason for the discrepancies.

Our measurements probe the small-scale cutoff in
the power ($\num{2e-2} \lesssim k \lesssim 0.1$) in all redshift bins
with a typical precision of \(5-15\%\).  The position of this cutoff
is still at far larger scales then the expected cutoff due to the
spectroscopic resolution of our data and the observed cutoff in
the power results from thermal and/or pressure broadening of the
absorption lines (or e.g. warm/fuzzy dark matter).
In the next subsection we also compare to
previous high-resolution measurements to make sure our measurement
agrees with existing data on small scales as well.

\subsection{Comparison to Previous High-Resolution Measurements} \label{results-compare-old-highres}

\begin{figure*}
\centering
\plotone{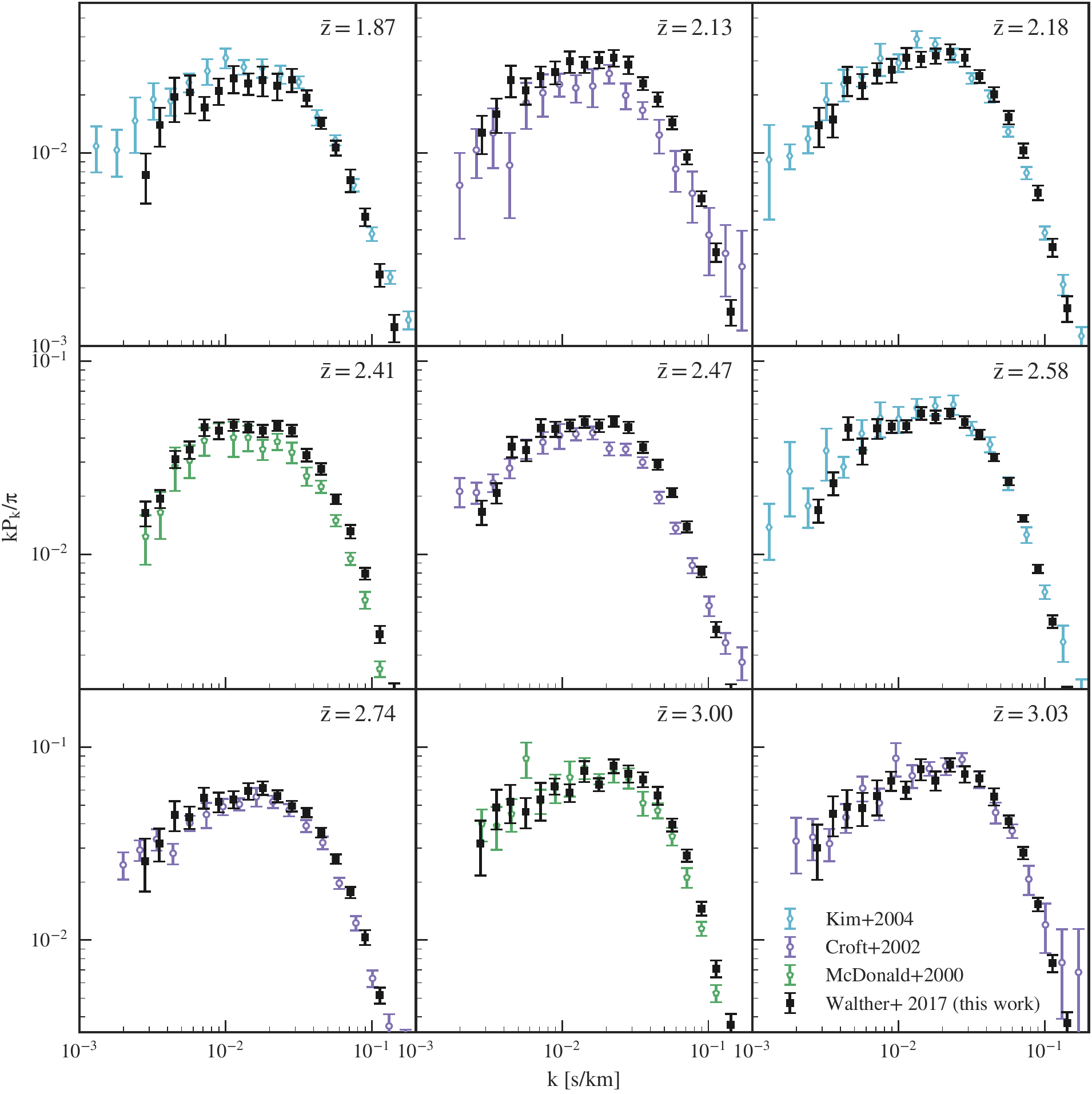}
\caption{Our new measurement of the power spectrum (black squares) compared to the existing measurements of \citet{mcdonald2000ObservedProbabilityDistribution}, \citet{croft2002PreciseMeasurementMatter} and \citet{kim2004powerspectrumflux}.
Our measurement has been interpolated between the 2 neighboring redshift bins to the same mean redshift as the other datasets.
The old measurements by \citet{mcdonald2000ObservedProbabilityDistribution} and \citet{croft2002PreciseMeasurementMatter} have been rescaled by \(\bar{F}^2\) as they were obtained on the flux itself instead of the flux contrast \(\delta_F\) which is the cause for different overall normalizations in some bins which is the cause for different overall normalizations in some bins.
The \citet{croft2002PreciseMeasurementMatter} measurement has been rescaled by an additional factor of two to match the Power spectrum normalization convention we use in this paper.
It is worth to remark that the older measurements were also performed in wider redshift bins (\(0.3\lesssim \Delta z \lesssim 0.6\)).
Also \citet{kim2004powerspectrumflux} performed their measurement on a subset of our dataset.
Notice that the approaches to noise, metal and resolution correction vary between all 4 datasets as well.
}\label{fig:powerspec_old_highres}
\end{figure*}

Previous power spectrum measurements based on high-resolution data were obtained by different groups using redshift bins of different size and location.
They also differ in the Fourier normalization convention used (leading to factors of 2 between some measurements), the field of which the power is computed (flux contrast or transmission leading to additional factors of \(\bar{F}^2\) in the power) as well as whether metals were masked and noise was subtracted.
We therefore show comparisons at the quoted redshifts for the old high-resolution measurements and linearly interpolate our results to those redshifts and renormalize the different measurements to the power spectrum convention that we use
(see \autoref{sec:app_conventions}).

For authors that chose to study the \(F\) field instead of \(\delta_F\), there is ambiguity of the mean flux. While the mean flux of the IGM is clearly the same, the mean flux of the data is sensitive to where the continuum is placed. It is well known that hand continuum fits to high-resolution spectra are biased low \citep{faucher-giguere2008EvolutionIntergalacticOpacity}, and this systematic effect is a bigger issue at higher redshift. If one measures the power spectrum
of \(F\), and the continuum fits are biased low, then the power will be biased high. \citet{mcdonald2000ObservedProbabilityDistribution} provide measurements of their mean flux, and we can therefore easily correct this effect, whereas \citet{croft2002PreciseMeasurementMatter} does not. Thus a direct comparison to Croft's measurements is not straightforward, but we do our best by simply assuming their continua are unbiased and multiplying in the mean flux of the IGM measured by \citep{becker2013refinedmeasurementmean} at the respective redshifts of their measurements.
The differences between power spectrum conventions clearly limit the precision at which our measurements will agree with previous work.

The comparison between high-resolution measurements is shown in \autoref{fig:powerspec_old_highres}.
While overall agreement is good considering the different approaches, some comparisons show disagreement.
The strongest mismatch is with \citep{croft2002PreciseMeasurementMatter} at \(z=2.13\) on all scales. At similar redshifts we agree with
  both \citet{kim2004powerspectrumflux} and \citetalias{palanque-delabrouille2013onedimensionalLy}  which hints toward an incorrect mean flux or a problem with the \citet{croft2002PreciseMeasurementMatter} measurement at this redshift (which is not part of their fiducial sample).
  We can also see that the shape of the different measurements on scales smaller (larger \(k\)) than the cutoff matches between the four high-resolution datasets for scales \(0.01\lesssim k\lesssim0.08\).  On smaller scales the difference in treatment of metals and S/N of the datasets as well as removal of noise power can probably explain the deviations between these measurements.
  We do note that our \(z=3.0\) bin exhibits a cutoff at slightly smaller scales (larger \(k\)) than the one of \citet{mcdonald2000ObservedProbabilityDistribution}, but agree with \citet{croft2002PreciseMeasurementMatter} at essentially the same redshift. This shows that there are clear limitations of comparisons to the previous measurements from high-resolution datasets due to the different conventions

To summarize, we find reasonable agreement between our measurements and previous analyses. We will use our new result for parameter estimations in future works.

 \section{Summary}
\label{sec:conclusions}

In this work we presented a new measurement of the \ac{Lya} forest power spectrum at \(1.8\leq z\leq3.4\) from archival high-resolution spectra obtained with the UVES and HIRES spectrographs. The pathlength of \(\sim \SI{20}{\com\giga\parsec}\) covered by this dataset (see \autoref{fig:metal-mask-lines}) is several times larger than the previous measurements \citet{mcdonald2000ObservedProbabilityDistribution}, \citet{croft2002PreciseMeasurementMatter}, and \citet{kim2004powerspectrumflux}. This allows us to measure the small scale cutoff in the power spectrum and its redshift evolution  with unprecedented precision.

We developed a custom pipeline to accurately measure the power spectrum
and its uncertainty, which fully corrects finite resolution and noise.
Some regions of quasar spectra must be masked because of missing data,
DLAs, and metal absorption line contaminants, which we identify and
mask using several methods.  If left uncorrected, this masking alters
the shape of the power spectrum, particularly on the small-scales
(high-$k$) of interest for studying the thermal state of the IGM. To
obtain unbiased estimates of the power spectrum and its associated
noise, we adopt a forward modeling approach. We post-process a DM
simulation (see \autoref{sec:dm-skewers} and \autoref{sec:fwmodel})
and generate a grid of different \ac{Lya} forest models with the same
noise and resolution as our data. The same masks are applied to these
mock spectra and we use a custom emulator (\autoref{sec:emulator}) to perform MCMC fits (\autoref{sec:param_estim}) to our
measurements. These models are sufficiently flexible that they provide
a good fit to the data, although the resulting parameters are not
physically meaningful. These model fits are then used to correct our
power spectrum and its covariance for the impact of masking (\autoref{sec:raw-measurement}).  Our
analysis shows that metal line contaminants significantly alter the
shape of the raw power spectrum on small-scales $k > \SI{0.1}{\skm}$.
Although our masking mitigates the effect of this contamination, we
nevertheless restrict further analysis of the power spectrum to $k<
\SI{0.1}{\skm}$. Our power spectrum measurements in 9 redshift bins covering
$1.8 < z < 3.4$ are presented in \autoref{sec:dataproducts}.

We compared our new measurements to previous results from both
low-/medium (\autoref{sec:final-measurement}) and high-resolution
(\autoref{results-compare-old-highres}) spectrographs.  Our
measurements agree well with the BOSS power spectrum \citepalias{palanque-delabrouille2013onedimensionalLy} for the low wavenumbers $k < \SI{0.02}{\skm}$ probed
by that low-resolution dataset. Given the extremely high $\sim 2\%$
precision of the BOSS study, we consider this an important validation
of our approach. Our measurements significantly disagree with the
recent study of \citep{irsic2016Lymanalphaforestpower} based on the medium-resolution XQ-100
dataset. This disagreement is restricted to redshift $z=3.2$ and
$z=3.4$ and is present only for the higher $k\gtrsim \SI{0.02}{\skm}$ modes. Given
the direction of the discrepancy, and the fact that only the highest
$k$ modes are affected, we argue that the disagreement most likely
results from over-correcting the effect of spectral resolution on the
power spectrum, which can ultimately be attributed to improper
characterization of the X-SHOOTER spectrograph's resolution. Comparing
our results to previous high-resolution efforts \citep{mcdonald2000ObservedProbabilityDistribution,croft2002PreciseMeasurementMatter,kim2004powerspectrumflux}, we
mostly find good agreement, although some combinations of dataset and redshift bin are discrepant
at the \(10-50\%\) level.  We do not believe these differences are a
significant source of concern, as they likely arise from the
challenges in comparing measurements covering significantly different
redshift ranges, adopting different mean flux normalization conventions (\autoref{sec:app_conventions}), and other systematics that may have plagued previous work.

Combining our precise small-scale (high-$k$) measurement with the larger scale power measured by \citetalias{palanque-delabrouille2013onedimensionalLy} and \citet{irsic2016Lymanalphaforestpower}
results in a powerful new dataset for placing percent level constraints on the thermal evolution of the IGM, as well as
cosmological parameters. In particular, the greatly improved high-$k$ precision enabled by our work will help break
degeneracies between the two, and enable improved constraints on neutrino masses,
as well alternatives to the CDM paradigm such as warm \citep{viel2013Warmdarkmatter,Irsic2017Newconstraints} or fuzzy dark matter  \citep{Hui2017Ultralightscalars,irsic2017Firstconstraintsfuzzy}.
In a companion paper we will compare these measurements to a grid of hydrodynamical simulations to place
precision constraints on the thermal evolution of the IGM, and search for the thermal signature of HeII reionization.
As our work only covers the redshift range $1.8 < z < 3.4$, future efforts should focus on filling in the gaps at
both lower $z\lesssim 1.8$ at $3.4 \lesssim z \lesssim 6$, which would enable studies of thermal state of the IGM from just
after hydrogen reionization until the present.

\section*{acknowledgments}
We thank Martin White for providing the collisionless simulation used for our models.
We'd like to thank the anonymous referee for useful comments.

We thank the members of the ENIGMA group\footnote{\url{http://enigma.physics.ucsb.edu/}} at the Max Planck Institute for Astronomy (MPIA) and University of California Santa Barbara (UCSB) for insightful suggestions  and  discussions.

Joseph  F.  Hennawi  acknowledges  generous support from the Alexander von Humboldt foundation in the context of the Sofja Kovalevskaja Award. The Humboldt foundation is funded by the German Federal Ministry for Education and Research.

Some data presented in this work ere obtained from the Keck Observatory Database of Ionized Absorbers toward QSOs (KODIAQ),
which was funded through NASA ADAP grants NNX10AE84G and NNx16AF52G along with NSF award number 1516777.

This research used resources of the National Energy Research Scientific Computing Center (NERSC), which is supported by the Office of Science of the U.S. Department of Energy under Contract no. DE-AC02-05CH11231.

We would like to thank ESO for making the ESO data archive publicly available.

\bibliography{powerspec}

 \appendix
 \section{Impacts of seeing on power spectrum measurements}
 \label{sec:reseffects}

For slit spectrographs, it is very difficult to know the exact resolution because it depends on the seeing, and also the resolution can vary at the $\sim 10\%$ across the echelle orders, which is typically never carefully quantified or taken into account \citep[e.g. the X-SHOOTER pipeline user manual][shows variations in the slit resolution with wavelength at about this level]{modigliani2017XshooterPipelineUser}.
Note that the change of resolution with seeing is not a problem for fiber spectrographs (such as e.g. BOSS) that allow measurements of the resolution of the science data on sky fibers.
Assuming the same resolution for each object (whereas the objects actually have different resolutions) generally will also increase the weight of higher resolution data in the power spectrum averages.
This is because higher resolution data has a smaller scale (higher \(k\)) cutoff. Therefore, a higher power by a factor that is exponential in k as well as in R would be measured.
When performing the mean over all objects the higher resolution objects (which now have been overcorrected) will bias the power estimate as \(\langle \exp(k^2R^2)\rangle > \exp(k^2\langle R\rangle^2)\) (if \(R\) is not constant throughout the dataset).
As explained in \citet{yeche2017Constraintsneutrinomasses} this problem can be weakened by measuring the seeing from the data (e.g. by measuring the width of the object in the spacial direction) and correcting each spectrum using its correct seeing.

A similar biasing due to mixing different resolutions can appear already during data reduction because co-adding overlapping echelle orders as well as different observations will give higher weight to data with better seeing as this typically has higher S/N as well.
If resolution varies strongly between echelle orders or observations, the final co-added spectrum might be dominated by the best resolution obtained.
Therefore, a knowledge of the spectrograph resolution for each individual object on the \(<10\%\) level as well as an individual resolution correction for each object are needed to provide an accurate measurement.

\section{Slit resolution of the X-SHOOTER spectrograph}
\label{sec:slitres}

  \begin{figure*}
  \plotone{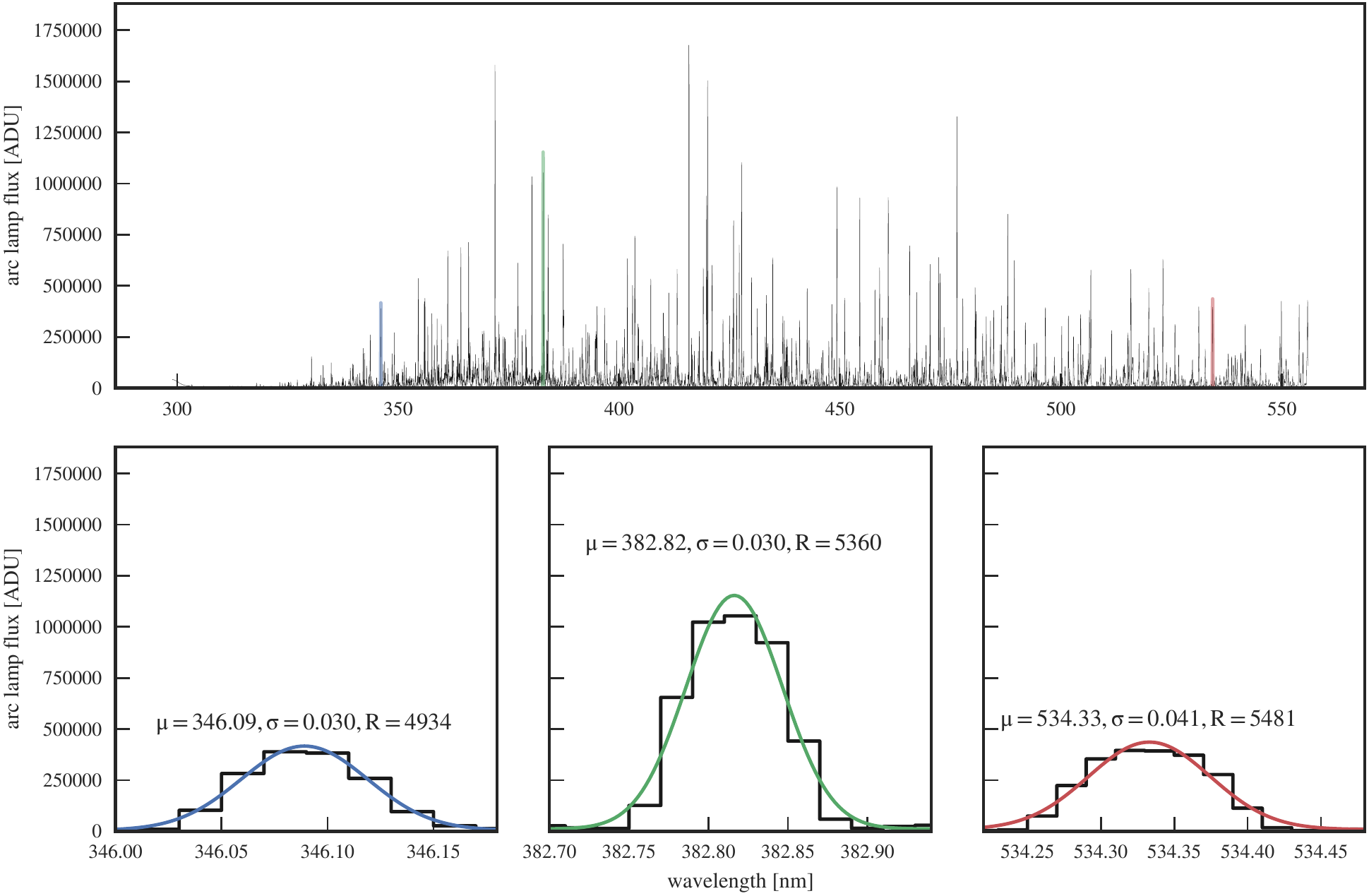}
   \caption{A reduced slit-arc spectrum of the UVB channel with the same slit width and binning as the XQ-100 survey and reduced with the \texttt{xsh\textunderscore{}scired\textunderscore{}slit\textunderscore{}stare} recipe of the ESO X-Shooter Pipeline using the same calibrations as for science data reductions (except for the response curves and disabling sky-subtraction). The full arc spectrum is shown in the top panel, colored lines indicate the position of zoom-ins in the bottom panel.
The bottom panels also show Gaussian fits (colored lines) to three of the arc lines with the best-fit parameters (mean \(\mu\), standard deviation \(\sigma\), both in nm) as well as the resulting resolving power \(R\) printed as text. We can clearly see that the resolution of each of those fits is exceeding the nominal value (taken from the XQ-100 data release paper) of \(R=4350\) and is varying over the spectrum.}
   \label{fig:slitres}
 \end{figure*}

 The quoted resolution in the X-SHOOTER manual was originally 5100 for the same slit/arm combination and was changed to 4260 during a recalibration run in 2011 (so before XQ-100 data was taken).
 The full reason for this change in value between calibrations is unclear to the authors.
 However, the X-SHOOTER Pipeline manual \citep{modigliani2017XshooterPipelineUser} shows values more consistent with the higher resolution.
 The manual also claims an underestimation of resolution by the pipeline for the 1x1 binning in the UVB arm.
 Additionally, the reduction QC plots in the \href{http://archive.eso.org/qc1/qc1_cgi?action=qc1_plot_table&table=xshooter_wavecal}{ESO archive} for 1x1 binning and 1x2 binning (which XQ-100 used) for all measurements between 2012 and 2014 shows \(\sim20\%\) higher values for the pipeline resolution values with the 1x2 binning.
  We contacted ESO about this issue and found out that the difference in estimated resolution for different binnings is still an open problem.
  A determination of the instruments' slit resolution to the accuracy needed for small scale power analyses is therefore not available.

\begin{figure}
  \plotone{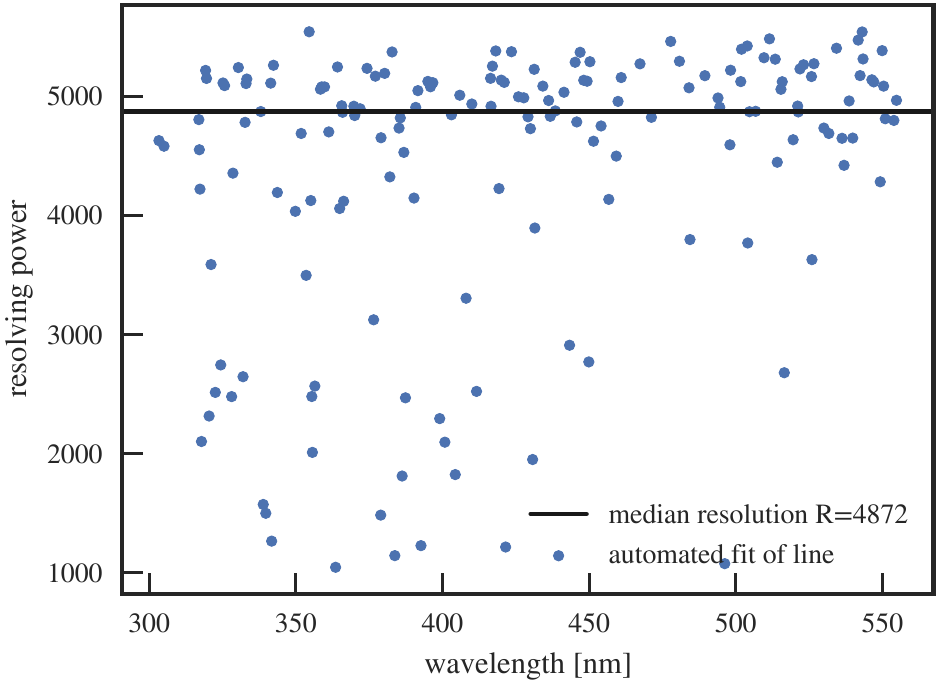}
   \caption{Full slit resolution with respect to wavelength for the X-SHOOTER spectrograph estimated from the spectrum in \autoref{fig:slitres}. We show the resolution for a quick fit of individual lines in the arc spectrum based on the line list provided by ESO. Obviously unphysical or unconverged fits have been removed. As fits to blends have not been removed, spurious low-resolution fits are included inside the figure.}
   \label{fig:slitres_of_wave}
 \end{figure}

 This motivated us to estimate the X-SHOOTER resolving power for the UVB arm and the configuration used in the XQ-100 dataset (\(\SI{0.1}{\arcsecond}\) slit width and a \(1\times2\) binning) from a slit-arc spectrum (taken at 2012-05-20T17:06:41.424) reduced in the same way as one would do for science data using the ESO Pipeline recipe \texttt{xsh\textunderscore{}scired\textunderscore{}slit\textunderscore{}stare}.
 These frames are taken on a regular basis using the same slits used for science targets.
 A normal reduction process (at least with the ESO pipeline) does not need those frames as the wavelength calibration is performed using pinhole arcs which cannot be used to determine the resolution of science data.
 In \autoref{fig:slitres} we show the reduced slit-arc spectrum as well as zoom-ins to three random, non-blended lines in different parts of the spectrum.
 We fitted the lines with Gaussian profiles which show that the resolution of X-SHOOTER in the UVB arm  a) varies within the arm by at least\(\sim 10\%\) and b) can be \(\sim 25\%\) higher than the value quoted in the XQ-100 data release paper \citep{lopez2016XQ100legacysurvey}.
 We also performed a quick automated fit for all lines in the arc line list individually without checking for line blends and other kinds of contamination.
 These might make some fits broader than a single line leading to determining a lower resolution.
 The resulting resolution with respect to wavelength is shown in \autoref{fig:slitres_of_wave}.
 We do not observe a clear trend with wavelength and note that the bulk of the distribution agrees with \(R\sim 5000\).

 These tests are in basic agreement with the QC plots in the ESO archive as well as the pipeline manual.
 We therefore assume that the resolution values given in the automatic QC plots are right and use their approximate median of \(R=5350\) when performing further tests of the XQ-100 results.
 If one only determines the seeing and estimates spectral resolution by combining the seeing with the slit resolution quoted on the ESO website or X-SHOOTER manual one might therefore under-determine the true resolution of the spectra.
 In principle one should be able to obtain the slit resolution from sky-lines in the science observations as well (albeit with less available lines and worse signal to noise than for the arc spectra) as motions in the atmosphere are slower than the \(\sim \SI{1}{\kms}\) resolution accuracy we want to obtain.

 \section{Normalization Conventions for the Power Spectrum}
 \label{sec:app_conventions}
 While \citet{kim2004powerspectrumflux} as well as the SDSS/BOSS/XQ-100 measurements already used the same normalization as we do, \citet{croft2002PreciseMeasurementMatter} and \citet{mcdonald2000ObservedProbabilityDistribution} measure the power of the transmission \(F\) instead of \(\delta_F\).
This leads to an additional factor \(\bar{F}^2\) between their measurements and the more recent ones.
The \citet{croft2002PreciseMeasurementMatter} measurement also has a different normalization convention by a factor 2 compared to \citet{mcdonald2000ObservedProbabilityDistribution} which we corrected for.
The former also does not provide a measurement of the mean transmission of their sample.
Therefore we renormalized the \citet{mcdonald2000ObservedProbabilityDistribution} measurement using the provided mean transmission of their sample and rescaled the \citet{croft2002PreciseMeasurementMatter} with the external mean transmission measurement by \citet{becker2013refinedmeasurementmean}.

\section{Data Products} \label{sec:dataproducts}
Truncated data tables showing our high-resolution measurement at \(z=2.8\) are shown in \autoref{tab:powerspec_final} (including our metal masking approach) and \autoref{tab:powerspec_final_nometal} (without metal masking).
We also show the first and last column of the correlation matrix at \(z=2.8\) in \autoref{tab:correlation_final}.
Note that the correlation matrix is based on the best-fitting DM only simulation.
It is therefore a model dependent quantity although the agreement of the fit with the data is good.
Thus, {while it should give a good representation of data correlations}, for fitting models to the measured power spectra you might want to estimate the correlation matrix from the actual model fitted in the analysis {to be fully independent of our modeling.}
The full tables including all redshifts and \(k\)-bins will be available in machine readable form.
{Masked spectra (with and without enabled metal masking) can be obtained from the Zenodo upload under \citet{walther_michael_2017_1041022}\footnote{\href{https://zenodo.org/record/1041022}{https://zenodo.org/record/1041022}}. Random samples from our \(f_\mathrm{window}\) chain can be found therein as well.}

\begin{deluxetable}{cccccc}
\tabletypesize{\footnotesize}
\tablecolumns{6}
\tablewidth{0pt}
\tablecaption{Priors used for each fitting parameter}
\tablehead{
\vspace{-0.15cm}&\colhead{type of}&lower&upper&&\\
\colhead{Parameter}\vspace{-0.15cm}&&&&\colhead{\(\mu\)}&\colhead{\(\sigma\)}\\
&\colhead{prior}&\colhead{limit}&\colhead{limit}&&
}
\startdata
\(\log(T_0/\si{K})\)&flat&3.48&4.30&&\\
\(\gamma\)&flat&0.5&2.1&&\\
\(\log(\lambda_J/\si{\ckpc})\)&flat&1.34&2.18&&\\
\(f_\mathSiIII\)& Gaussian & -0.002 & 0.018 & 0.008 & 0.001\\
\(\bar{F}(z=1.8)\)& Gaussian & 0.871 & 0.931 & 0.901 & 0.006\\
\(\bar{F}(z=2.0)\)& Gaussian & 0.785 & 0.976 & 0.881 & 0.019\\
\(\bar{F}(z=2.2)\)& Gaussian & 0.818 & 0.921 & 0.869 & 0.010\\
\(\bar{F}(z=2.4)\)& Gaussian & 0.766 & 0.859 & 0.812 & 0.009\\
\(\bar{F}(z=2.6)\)& Gaussian & 0.723 & 0.813 & 0.768 & 0.009\\
\(\bar{F}(z=2.8)\)& Gaussian & 0.683 & 0.771 & 0.727 & 0.009\\
\(\bar{F}(z=3.0)\)& Gaussian & 0.640 & 0.724 & 0.682 & 0.008\\
\(\bar{F}(z=3.2)\)& Gaussian & 0.581 & 0.661 & 0.621 & 0.008\\
\(\bar{F}(z=3.4)\)& Gaussian & 0.528 & 0.602 & 0.565 & 0.007\\
 \label{tab:priors}
\enddata
\tablecomments{priors in \(\bar{F}\) are based on \citet{kirkman2005HIopacityintergalactic} for \(z=1.8\), \citet{faucher-giguere2008DirectPrecisionMeasurement} for \(z=2\) and \citet{becker2013refinedmeasurementmean} for the higher redshifts}
\end{deluxetable}

\begin{deluxetable}{cccc}
\tabletypesize{\footnotesize}
\tablecolumns{4}
\tablewidth{0pt}
\tablecaption{Measured flux power spectrum at \(z=2.8\) after masking metals and removing the window function due to masking. Note that the full table that also includes the other redshifts will be available in the electronic edition of the journal.}
\tablehead{
  \vspace{-0.15cm}&\colhead{\(\bar{k}\)}&&\\
  \colhead{\(\bar{z}\)}\vspace{-0.15cm}&& \colhead{\(k P_k \pi^{-1}\)}&  \colhead{\(\sigma_{k P_k \pi^{-1}}\)}\\
  &\colhead{\si{\skm}}&&
}
\startdata
\input{ps_table_metalmasked}
\enddata
\label{tab:powerspec_final}

\end{deluxetable}

\begin{deluxetable}{cccc}
\tabletypesize{\footnotesize}
\tablecolumns{4}
\tablewidth{0pt}
\tablecaption{Measured flux power spectrum at \(z=2.8\) without masking of metals and after removing the window function due to masking. Note that the full table that also includes the other redshifts will be available in the electronic edition of the journal.}
\tablehead{
  \vspace{-0.15cm}&\colhead{\(\bar{k}\)}&&\\
  \colhead{\(\bar{z}\)}\vspace{-0.15cm}&& \colhead{\(k P_k \pi^{-1}\)}&  \colhead{\(\sigma_{k P_k \pi^{-1}}\)}\\
  &\colhead{\si{\skm}}&&
}
\startdata
\input{ps_table}
\enddata
 \label{tab:powerspec_final_nometal}
\end{deluxetable}

\begin{deluxetable}{ccccc}
\tabletypesize{\footnotesize}
\tablecolumns{5}
\tablewidth{0pt}
\tablecaption{Correlation matrix at \(z=2.8\) for the measurement in \autoref{tab:powerspec_final}. Note that the full table that also includes the other redshifts and matrix elements will be available in the electronic edition of the journal.}
\tablehead{
  \vspace{-0.15cm}&\colhead{\(\bar{k}\)}&&& \\
  \colhead{\(\bar{z}\)}\vspace{-0.15cm}&& \colhead{\(R_{1,j}\)}& \colhead{\(\dots\)} &  \colhead{\(R_{n,j}\)} \\
  &\colhead{\si{\skm}}&&& \\
}
\startdata
 \input{corr_table_metalmasked}
\enddata
\label{tab:correlation_final}

\end{deluxetable}

\begin{acronym}
 \acro{DLA}{damped \acs{Lya} absorption system}
 \acro{UVB}{ultraviolet background}
 \acro{LLS}{Lyman Limit System}
 \acro{pLLS}{partial \ac{LLS}}
 \acro{Lya}[Ly\textalpha]{Lyman Alpha}
 \acro{DM}{dark matter}
 \acro{UVES}{Ultraviolet and Visual Echelle Spectrograph}
 \acro{HIRES}{High Resolution Echelle Spectrometer}
 \acro{KODIAQ}{Keck Observatory Database of Ionized Absorption toward Quasars}
 \acro{DR}{Data Release}
 \acro{VLT}{Very Large Telescope}
 \acro{SDSS}{Sloan Digital Sky Survey}
 \acro{BOSS}{Baryon Oscillation Spectroscopic Survey}
 \acro{IGM}{intergalactic medium}
 \acro{MCMC}{Markov Chain Monte Carlo}
 \acro{1DPS}{1d flux power spectrum}
 \acro{FGPA}{fluctuating Gunn-Peterson absorption}
 \acro{PDF}{probability density function}
 \acro{PCA}{principal component analysis}
 \acro{AMR}{adaptive mesh refinement}
 \acro{PM}{particle-mesh}
 \acro{BAL}{broad absorption line}
 \acro{EoR}{Epoch of Reionization}
 \acro{PPM}{piecewise parabolic method}
 \acro{UV}{ultraviolet}
 \acro{HST}{Hubble Space Telescope}
 \acro{COS}{Cosmic Origins Spectrograph}
 \acro{QSO}{quasar}

\end{acronym}

\end{document}

%% file: uves_table.tex
HE1341\(-\)1020 & 2.137 & 58.1 \\
Q0122\(-\)380 & 2.192 & 56.4 \\
PKS1448\(-\)232 & 2.222 & 57.4 \\
PKS0237\(-\)23 & 2.224 & 102.4 \\
HE0001\(-\)2340 & 2.278 & 65.9 \\
Q0109\(-\)3518 & 2.406 & 70.0 \\
HE1122\(-\)1648 & 2.407 & 171.6 \\
HE2217\(-\)2818 & 2.414 & 93.6 \\
Q0329\(-\)385 & 2.437 & 58.4 \\
HE1158\(-\)1843 & 2.459 & 66.7 \\
Q2206\(-\)1958 & 2.567 & 74.5 \\
Q1232+0815 & 2.575 & 45.8 \\
HE1347\(-\)2457 & 2.615 & 62.0 \\
HS1140+2711 & 2.628 & 88.9 \\
Q0453\(-\)423 & 2.663 & 77.6 \\
PKS0329\(-\)255 & 2.705 & 48.0 \\
Q1151+068 & 2.758 & 49.1 \\
Q0002\(-\)422 & 2.768 & 75.0 \\
HE0151\(-\)4326 & 2.787 & 98.1 \\
Q0913+0715 & 2.788 & 54.4 \\
Q1409+095 & 2.843 & 24.7 \\
HE2347\(-\)4342 & 2.886 & 152.3 \\
Q1223+178 & 2.955 & 33.4 \\
Q0216+08 & 2.996 & 36.8 \\
HE2243\(-\)6031 & 3.011 & 118.8 \\
CTQ247 & 3.026 & 69.1 \\
HE0940\(-\)1050 & 3.089 & 69.6 \\
Q0420\(-\)388 & 3.120 & 116.2 \\
CTQ460 & 3.141 & 40.9 \\
Q2139\(-\)4434 & 3.208 & 31.2 \\
Q0347\(-\)3819 & 3.229 & 83.9 \\
PKS2126\(-\)158 & 3.285 & 63.6 \\
Q1209+0919 & 3.291 & 30.2 \\
Q0055\(-\)269 & 3.665 & 75.7 \\
Q1249\(-\)0159 & 3.668 & 69.7 \\
Q1621\(-\)0042 & 3.708 & 77.7 \\
Q1317\(-\)0507 & 3.719 & 42.0 \\
PKS2000\(-\)330 & 3.786 & 150.8

%% file: kodiaq_table.tex
J122824+312837 & 2.200 & 87.3 \\
J110610+640009 & 2.203 & 58.5 \\
J162645+642655 & 2.320 & 103.7 \\
J141906+592312 & 2.321 & 36.7 \\
J005814+011530\tablenotemark{c} & 2.495 & 36.2 \\ 
J162548+264658\tablenotemark{c} & 2.518 & 43.9 \\ 
J121117+042222 & 2.526 & 33.6 \\
J101723\(-\)204658 & 2.545 & 70.3 \\
J234628+124859 & 2.573 & 75.1 \\
J101155+294141\tablenotemark{a} & 2.620 & 129.9 \\
J082107+310751 & 2.625 & 64.0 \\
J121930+494052 & 2.633 & 90.3 \\
J143500+535953 & 2.635 & 65.0 \\
J144453+291905 & 2.669 & 133.7 \\
J081240+320808 & 2.712 & 48.8 \\
J014516\(-\)094517A & 2.730 & 76.8 \\
J170100+641209\tablenotemark{c} & 2.735 & 81.8 \\ 
J155152+191104 & 2.830 & 30.2 \\
J012156+144820 & 2.870 & 54.5 \\ 
Q0805+046\tablenotemark{b} & 2.877 & 26.8 \\
J143316+313126 & 2.940 & 53.8 \\
J134544+262506 & 2.941 & 34.7 \\
J073621+651313 & 3.038 & 25.7 \\
J194455+770552\tablenotemark{a} & 3.051 & 30.4 \\
J120917+113830 & 3.105 & 31.4 \\
J114308+345222\tablenotemark{a} & 3.146 & 31.9 \\
J102009+104002 & 3.168 & 35.9 \\
J1201+0116\tablenotemark{b} & 3.233 & 30.1 \\
J080117+521034\tablenotemark{a} & 3.236 & 43.2 \\
J095852+120245\tablenotemark{a} & 3.298 & 44.8 \\
J025905+001126 & 3.365 & 26.3 \\
Q2355+0108\tablenotemark{b} & 3.400 & 58.3 \\
J173352+540030 & 3.425 & 57.3 \\
J144516+095836\tablenotemark{a} & 3.530 & 24.6 \\
J142438+225600\tablenotemark{a} & 3.630 & 29.3 \\
J193957\(-\)100241 & 3.787 & 65.5 

%% file: metallist.tex
\ion{O}{6}\tablenotemark{a}  & 1031.9261        & \ion{Si}{4}\tablenotemark{a} & 1402.770 \\
\ion{C}{2}         & 1036.3367        & \ion{Si}{2}        & 1526.7066 \\
\ion{O}{6}         & 1037.6167        & \ion{C}{4}\tablenotemark{a}  & 1548.195 \\
\ion{N}{2}         & 1083.990         & \ion{C}{4}\tablenotemark{a}  & 1550.770\\
\ion{Fe}{3}        & 1122.526         & \ion{Fe}{2}        & 1608.4511\\
\ion{Fe}{2}        & 1144.9379        & \ion{Al}{2}        & 1670.7874\\
\ion{Si}{2}        & 1190.4158        & \ion{Al}{3}        & 1854.7164\\
\ion{Si}{2}        & 1193.2897        & \ion{Al}{3}        & 1862.7895\\
\ion{N}{1}         & 1200.7098        & \ion{Fe}{2}        & 2344.214\\
\ion{Si}{3}\tablenotemark{a} & 1206.500         & \ion{Fe}{2}        & 2374.4612\\
\ion{N}{5}         & 1238.821         & \ion{Fe}{2}        & 2382.765\\
\ion{N}{5}         & 1242.804         & \ion{Fe}{2}        & 2586.6500\\
\ion{Si}{2}\tablenotemark{a} & 1260.4221        & \ion{Fe}{2}        & 2600.1729\\
\ion{O}{1}         & 1302.1685        & \ion{Mg}{2}        & 2796.352\\
\ion{Si}{2}        & 1304.3702        & \ion{Mg}{2}        & 2803.531\\
\ion{C}{2}         & 1334.5323        & \ion{Mg}{1}        & 2852.9642\\
\ion{C}{2}*        & 1335.7077        & \ion{Ca}{1}        & 3934.777\\
\ion{Si}{4}\tablenotemark{a} & 1393.755         & \ion{Ca}{1}        & 3969.591

%% file: ps_table_metalmasked.tex
2.795 & 0.002803 & 0.03103 & 0.009415 \\
2.795 & 0.003499 & 0.03522 & 0.006892 \\
2.795 & 0.004479 & 0.04468 & 0.008296 \\
2.795 & 0.005632 & 0.04717 & 0.006015 \\
2.795 & 0.00708 & 0.05932 & 0.007473 \\
2.795 & 0.008945 & 0.05596 & 0.006969 \\
2.795 & 0.0113 & 0.05733 & 0.005735 \\
2.795 & 0.01425 & 0.06204 & 0.006429 \\
2.795 & 0.01794 & 0.06517 & 0.005423 \\
2.795 & 0.02259 & 0.05688 & 0.003983 \\
2.795 & 0.02838 & 0.05032 & 0.002825 \\
2.795 & 0.03573 & 0.04715 & 0.002449 \\
2.795 & 0.04501 & 0.03812 & 0.002463 \\
2.795 & 0.05666 & 0.02728 & 0.001632 \\
2.795 & 0.07132 & 0.01872 & 0.001295 \\
2.795 & 0.08978 & 0.01121 & 0.00106 \\
2.795 & 0.113 & 0.00548 & 0.0005469 \\
2.795 & 0.1423 & 0.002266 & 0.0003068 \\
2.795 & 0.1792 & 0.0009699 & 0.000172 \\
2.795 & 0.2255 & 0.0003949 & 9.197e-05 \\
2.795 & 0.2839 & 0.0002304 & 3.592e-05 \\
2.795 & 0.3574 & 0.0001749 & 2.872e-05 \\

%% file: ps_table.tex
2.797 & 0.002822 & 0.02772 & 0.007839 \\
2.797 & 0.003539 & 0.03315 & 0.007177 \\
2.797 & 0.004509 & 0.04198 & 0.006247 \\
2.797 & 0.005635 & 0.04608 & 0.005853 \\
2.797 & 0.007045 & 0.04969 & 0.006732 \\
2.797 & 0.00895 & 0.05159 & 0.005478 \\
2.797 & 0.01133 & 0.06545 & 0.006951 \\
2.797 & 0.0142 & 0.06722 & 0.006563 \\
2.797 & 0.01791 & 0.06762 & 0.004682 \\
2.797 & 0.02264 & 0.06177 & 0.005428 \\
2.797 & 0.02843 & 0.05868 & 0.003455 \\
2.797 & 0.03575 & 0.04829 & 0.003225 \\
2.797 & 0.045 & 0.04044 & 0.002322 \\
2.797 & 0.05662 & 0.03103 & 0.002069 \\
2.797 & 0.07131 & 0.02063 & 0.001187 \\
2.797 & 0.08978 & 0.01292 & 0.001051 \\
2.797 & 0.113 & 0.007671 & 0.0006593 \\
2.797 & 0.1423 & 0.004842 & 0.0005712 \\
2.797 & 0.1792 & 0.003066 & 0.0003734 \\
2.797 & 0.2255 & 0.001638 & 0.0002774 \\
2.797 & 0.2839 & 0.0008713 & 0.0001418 \\
2.797 & 0.3574 & 0.0005683 & 0.0001228 \\

%% file: corr_table_metalmasked.tex
2.8 & 0.002803 & 1 && -0.08482 \\
2.8 & 0.003499 & 0.1022 && 0.009749 \\
2.8 & 0.004479 & 0.14 && -0.05212 \\
2.8 & 0.005632 & 0.1838 && -0.02332 \\
2.8 & 0.00708 & 0.1645 && -0.002453 \\
2.8 & 0.008945 & 0.112 && 0.04121 \\
2.8 & 0.0113 & 0.03341 && -0.04957 \\
2.8 & 0.01425 & 0.07737 && -0.07281 \\
2.8 & 0.01794 & -0.001218 && 0.04444 \\
2.8 & 0.02259 & 0.05922 && 0.08884 \\
2.8 & 0.02838 & 0.04739 && 0.1806 \\
2.8 & 0.03573 & -0.08184 && 0.2035 \\
2.8 & 0.04501 & -0.1025 && 0.2783 \\
2.8 & 0.05666 & -0.08604 && 0.3239 \\
2.8 & 0.07132 & -0.09288 && 0.4122 \\
2.8 & 0.08978 & -0.1069 && 0.5116 \\
2.8 & 0.113 & -0.07494 && 0.4853 \\
2.8 & 0.1423 & -0.0566 && 0.4612 \\
2.8 & 0.1792 & -0.0352 && 0.3981 \\
2.8 & 0.2255 & -0.03103 && 0.3749 \\
2.8 & 0.2839 & -0.06502 && 0.6858 \\
2.8 & 0.3574 & -0.08482 && 1 \\